\documentclass[a4paper,oneside,final,notitlepage,onecolumn,12pt]{article}
\usepackage{amsfonts}
\usepackage{epsf}
\usepackage{graphicx}
\usepackage{amssymb,eso-pic}
\usepackage{latexsym}
\usepackage{tabularx}
\usepackage{amsxtra} 
\usepackage{hyperref}
\usepackage{t1enc}
\usepackage{amsmath,accents}
\usepackage{ifpdf,epsfig,array,amsmath,amssymb}

\usepackage{psfrag} 
\psfrag{na}{\footnotesize $n^a$}   
\psfrag{ta}{\footnotesize $\tau^a$}   
\psfrag{vna}{\footnotesize $\check{n}^a$}
\psfrag{vNa}{\footnotesize $\check{N}^a$}
\psfrag{scrW}{\footnotesize $\mycal{W}_{\rho_0}$}
\psfrag{t=t1}{\footnotesize $\Sigma_{\tau_1}$}
\psfrag{t=t2}{\footnotesize $\Sigma_{\tau_2}$}
\psfrag{r=all}{\footnotesize $\rho=const$}
\psfrag{hna}{\footnotesize $\hat{n}^a$}
\psfrag{tna}{\footnotesize $\tilde{n}^a$}
\psfrag{hab,Kab}{\footnotesize $h_{ab}, K_{ab}$}
\psfrag{thab,tKab}{\footnotesize $\tilde h_{ab}, \tilde K_{ab}$}
\psfrag{hhab,hKab}{\footnotesize $\hat h_{ab}, \hat K_{ab}$}
\psfrag{chab,cKab}{\footnotesize $\check{h}_{ab}, \check{K}_{ab}$}

\usepackage[normalem]{ulem}
\usepackage{color}
\definecolor{blue}{rgb}{0,0,1}
\definecolor{red}{rgb}{1,0,0}

\newcommand{\interior}[1]{\accentset{\smash{\raisebox{-0.12ex}{$\scriptstyle\circ$}}}{#1}\rule{0pt}{2.3ex}}

\DeclareFontFamily{OT1}{rsfs}{} \DeclareFontShape{OT1}{rsfs}{m}{n}{
<-7> rsfs5 <7-10> rsfs7 <10-> rsfs10}{}
\DeclareMathAlphabet{\mycal}{OT1}{rsfs}{m}{n}

\def\scri{{\mycal I}}
\def\scrip{\scri^{+}}%

\def\sc{{\hskip 3.5pt {{}^{{}^{{}_{{}_{\bowtie}}}}} \kern -8.pt{}}}  
\def\SC{{\hskip 3.5pt {{}^{{}^{{}^{{}_{{}_{\bowtie}}}}}} \kern -10.5pt{}}}

\DeclareMathAlphabet{\mathpzc}{OT1}{pzc}{m}{it}

\begin{document}

\newtheorem{theorem}{Theorem}[section]
\newtheorem{lemma}{Lemma}[section]
\newtheorem{proposition}{Proposition}[section]
\newtheorem{corollary}{Corollary}[section]
\newtheorem{conjecture}{Conjecture}[section]
\newtheorem{condition}{Condition}[section]
\newtheorem{example}{Example}[section]
\newtheorem{definition}{Definition}[section]
\newtheorem{remark}{Remark}[section]
\newtheorem{exercise}{Exercise}[section]
\newtheorem{axiom}{Axiom}[section]
\renewcommand{\theequation}{\thesection.\arabic{equation}} 

\author{{Istv\'an R\'acz}\,\thanks{ ~email: racz.istvan@wigner.mta.hu}  
\\ 
{Wigner RCP}\\  {H-1121 Budapest, Konkoly Thege Mikl\'os \'ut 29-33. Hungary
}}
\title{Cauchy problem as a two-surface based `geometrodynamics'\,\footnote{Dedicated to Zolt\'an Cseke on the occasion of his 70th Birthday.}} 

\maketitle

\begin{abstract}
Four-dimensional spacetimes foliated by a two-parameter family of homologous two-surfaces are considered in Einstein's theory of gravity. By combining a $1+(1+2)$ decomposition, the canonical form of the spacetime metric and a suitable specification of the conformal structure of the foliating two-surfaces a gauge fixing is introduced. It is shown that, in terms of the chosen geometrically distinguished variables, the  $1+3$ Hamiltonian and momentum constraints can be recast into the form of a parabolic equation and a first order symmetric hyperbolic system, respectively. Initial data to this system can be given on one of the two-surfaces foliating the three-dimensional initial data surface. The $1+3$ reduced Einstein's equations are also determined. 
By combining the $1+3$ momentum constraint with the reduced system of the secondary $1+2$ decomposition a mixed hyperbolic-hyperbolic system is formed. It is shown that solutions to this mixed hyperbolic-hyperbolic system are also solutions to the full set of Einstein's equations provided that the $1+3$ Hamiltonian constraint is solved on the initial data surface $\Sigma_0$ and the $1+2$ Hamiltonian and momentum type expressions vanish on a world-tube yielded by the Lie transport of one of the two-surfaces foliating $\Sigma_0$ along the time evolution vector field. Whenever the foliating two-surfaces are compact without boundary in the spacetime and a regular origin exists on the time-slices---this is the location where the foliating two-surfaces smoothly reduce to a point---it suffices to guarantee that the $1+3$ Hamiltonian constraint holds on the initial data surface.  
A short discussion on the use of the geometrically distinguished variables in identifying the degrees of freedom of gravity are also included.
\end{abstract} 

\date


\section{Introduction}\label{introduction}
\setcounter{equation}{0}
Since the fundamental discovery by Yvonne Choquet-Bruhat \cite{bruhat} there have been considerable developments on the hyperbolic formulations of various coupled Einstein--matter systems in general relativity (see e.g. \cite{helmut,helmut&alan,geroch,reula} and references therein). The main issue in all of these approaches is to show the existence and uniqueness of solutions to a suitably separated subset of the field equations (usually referred as the reduced set of evolution equations) and also to demonstrate that the rest of the equations (the constraints) will hold in the associated `domain of existence' provided they are satisfied on the initial data surface.  

\medskip

The separation of the field equations is usually done by combining it with a simultaneous gauge fixing in order to acquire further simplifications. To motivate the search for suitable gauge choices let us recall that in an arbitrary local coordinate system the Ricci tensor (see e.g.~(10.2.25) in \cite{wald}) take the form 
\begin{equation}\label{helmut_R}
R_{\mu\nu}=-\frac{1}{2}\,g^{\varepsilon\sigma}\left\{\partial_\varepsilon\partial_\sigma g_{\mu\nu}+\partial_\mu\partial_\nu g_{\varepsilon\sigma}-\partial_\varepsilon\partial_\nu g_{\mu\sigma}-\partial_\mu\partial_\sigma g_{\varepsilon\nu}\right\}+F_{\mu\nu}(g_{\lambda\kappa},\partial_{\gamma}g_{\lambda\kappa})
\end{equation} 
where $\partial_{\alpha}$ denotes the partial derivative operator with respect to local coordinates $x^\alpha$ and $F_{\mu\nu}$ can be given as $F'_{\mu\nu}/[\det(g)]^2$ where $F'_{\mu\nu}$ is a polynomial of degree eight in $g_{\varepsilon\sigma}$ and $\partial_{\gamma}g_{\varepsilon\sigma}$.
To see this, note that $F_{\mu\nu}$ is quadratic in the Christoffel symbols and the latter involve---besides a linear combination of the first derivatives of $g_{\mu\nu}$---the inverse metric $g^{\mu\nu}$ which can be given as $g'^{\mu\nu}/\det(g)$ , where $g'^{\mu\nu}$ can at best be determined as a polynomial of degree three the components $g_{\mu\nu}$.
These simple observations indicate that a suitable gauge choice could help not merely in reducing the complexity of metric inversion but also in simplifying the specific form of $F_{\mu\nu}$.  
In this paper, by investigating the Cauchy problem in smooth four-dimensional spacetimes which are smoothly foliated by a two-parameter family of homologous two-surfaces, a specific new gauge choice is proposed.

\medskip

Before proceeding it is important to emphasize that in general relativity various types of $1+(1+2)$ decompositions have already been applied. For example whenever the underlying spacetime admits non-null symmetries the methods proposed by Geroch in his seminal papers \cite{geroch2,geroch3} can be used. A different type of $1+(1+2)$ decomposition can be done by performing first a $1+3$ reduction based on a spacetime symmetry {\it a la} \cite{geroch2} which is succeeded then by an ADM type $1+2$ splitting (see e.g.~\cite{maeda,oliver}). 

\medskip

There are also remarkable developments based on appropriately tailored $1+2$ splittings of time level surfaces related either to the proof of the positivity of gravitational energy \cite{jj1,jj2,jj3} or to the existence of quasi-spherical metrics with prescribed scalar curvature on Cauchy surfaces \cite{bartnik,schwein1,schwein2}. Although the aims and techniques applied in \cite{jj1,jj2,jj3} and in \cite{bartnik,schwein1,schwein2} differ they are common in that in both approaches when the foliating two-surfaces are compact they are assumed not only to be topological two-spheres but the induced metric on them is also supposed to be conformal to that of a metric sphere. Interestingly, the Hamiltonian constraint becomes an elliptic equation in the setup used in \cite{jj1,jj2,jj3} whereas it is found to be parabolic in the approach applied in \cite{bartnik,schwein1,schwein2}. The main reason beyond these differences is that the constraint equations always comprise an underdetermined system. This, in turn, 
provides the freedom to choose the dependent variables according to the needs of the problem to be solved. Remarkably, the very same equation which is parabolic in \cite{bartnik,schwein1,schwein2}, when it is taken as a partial differential equation (PDE) for the three-dimensional lapse function, it can be seen to be elliptic if it is 
considered to be a PDE for the conformal factor relating the induced metric on the foliating spheres to that of the metric sphere.\,\footnote{Note that equation (11) of \cite{schwein1}, along with the determination of the coefficients $\bar A$ and $\bar B$ on page 55 can be used to verify this statement. In particular, while equation (11) of \cite{schwein1} is a parabolic PDE for the lapse $u$ the very same equation gets to be an elliptic PDE for the conformal factor, i.e.~for $v$. To see this note that $\partial_{\bar N}\bar H$ hides the apparently missing second $r$-derivative of $v$. (See also Subsection \ref{constraints2nd} below for further discussions in a more generic setup.)} 

\medskip

It is also important to keep in mind that the aforementioned investigations focused on the study of the constraint equations on a single time slice. Thereby if the gauge fixings used in these investigations had been applied in studying evolution one could have immediately realized that to retain the applied particular gauge fixing on the succeeding time level surfaces the four-dimensional shift vector had to evolve accordingly. This restriction comes as a price for the use of part of the four-dimensional diffeomorphism invariance to put the three-metric into a mathematically convenient form suiting to the arguments in \cite{jj1,jj2,jj3,bartnik,schwein1,schwein2}. As opposed to these approaches we prefer to use the canonical form of the metric (\ref{pref}) with vanishing four-dimensional shift vector thereby we have to use the most generic form of the three-metric on the succeeding time slices. Interestingly, for the gauge choice we made the specific form of the three-metric involves two variables in 
addition to the ones used in \cite{jj1,jj2,jj3,bartnik,schwein1,schwein2}. As it will be shown in Section \ref{final} these supplementary variables turn out to play an important role in identifying the true degrees of freedom of Einstein's theory of gravity.

\medskip

Concerning the more ambitious $1+(1+2)$ generic decompositions let us start by mentioning the remarkable results covered by a series of papers \cite{diverno1,diverno2,diverno3} based on some seminal visions by Stachel and d'Inverno. These papers, especially \cite{diverno1}, provide excellent conceptual perspectives of many of the basic issues. Note however, that there are also limitations concerning the applications of the pertinent results. The limitations originate partly in the complexity of the addressed problems and in that the scope of investigations were kept too wide, e.g.~by aiming to put all kinds of initial value problems of general relativity into a single universal framework. This could not allow to reach the roots of all the involved problems. More importantly, in \cite{diverno1,diverno2,diverno3} the investigations did not go beyond the derivation of some formal relations. In particular, a clear constructive 
introduction of the conformal structure or the study of the {solubility} of the involved constraint and evolutionary systems were left out from considerations in \cite{diverno1,diverno2,diverno3}. 

\medskip

Last but not least we would like to mention \cite{oliver&vince} containing some recent investigations in context of $1+(1+2)$ decompositions. The main concern of the authors there was to evaluate certain apparently singular expressions (arising in the conformally rescaled Einstein's equations), by making use of the L'Hopital rule, at $\scrip$. Thereby attention in \cite{oliver&vince} was restricted to a small neighborhood of future null infinity and no attempt was made to use generic $1+(1+2)$ decompositions in the Cauchy problem. 

\medskip

In this paper $1+(1+2)$ decompositions are applied in a generic setup. More specifically, the Cauchy problem is studied in Einstein's theory of gravity assuming that the base manifold is smoothly foliated by a two-parameter family of homologous two-surfaces. By making use of $1+(1+2)$ decompositions, the canonical form of the spacetime metric (\ref{pref}) \cite{chris,MOSM}, and a constructional definition of the conformal structure of the foliating two-surfaces a geometrically distinguished set of dependent variables are chosen. By making use of the associated gauge fixing the followings could be done: 
\begin{itemize}
\item[(1)] The Hamiltonian and momentum constraints are put into a system comprised by a parabolic equation and a first order symmetric hyperbolic system. 
\item[(2)] Provided that the foliating two-surfaces are orientable compact manifolds with no boundary in $M$ to this parabolic-hyperbolic system an initial value problem is set up. The pertinent initial values can be given on one of the two-surfaces foliating the three-dimensional initial data surface. 
\item[(3)] The reduced set of Einstein's equations is also determined. The first order symmetric hyperbolic part of the constraint equations, i.e.~the momentum constraint, and the $1+2$ reduced  equations---the latter are also manifestly strongly hyperbolic---can be used to set up a mixed hyperbolic-hyperbolic system to be solved. 
\item[(4)] It is shown that solutions to this mixed hyperbolic-hyperbolic system are also solutions to the full set of Einstein's equations provided that the $1+3$ Hamiltonian constraint is satisfied on the initial data surface $\Sigma_0$ and the $1+2$ Hamiltonian and momentum type expressions vanish on a world-tube yielded by the Lie transport of one of the two-surfaces foliating $\Sigma_0$ along the time evolution vector field. 
\end{itemize}

One of the most important new observations is that the $1+3$ constraints which---in the conventional setup are elliptic equations and they are solved by the conformal method \cite{wheeler, york, york0, fisher}---can be recast into the form a coupled parabolic-hyperbolic system to which an initial value problem can be applied whenever the two-surfaces foliating the spacetime manifold $M$ are orientable compact manifolds with no boundary in $M$. 
It appears to be equally important that a mixed hyperbolic-hyperbolic system of equations can be set up in solving the evolutionary part of the Cauchy problem in Einstein's theory of gravity. Note also that both of these new approaches appear to provide preferable solutions to the related issues. 

\medskip

Interestingly both the constraint and evolutionary equations can exclusively be expressed in terms of geometrically distinguished variables, along with their transversal derivatives. Each of these terms are defined on the foliating two-surfaces. This indicates that in the chosen topological setup the conventional Cauchy problem may also be viewed as a two-surface based `geometrodynamics'. 
 
\medskip

Remarkably, the constructive elements of the proposed new gauge fixing can also be used to flash lights on the true degrees of freedom of Einstein's theory of gravity. (For a more adequate dynamical determination of these degrees of freedom see \cite{racz_tdfd}.) 
 
\medskip

The paper is organized as follows. In Section \ref{prelim} the geometric and topological assumptions, fixing the applied kinematic setup, are introduced. Section \ref{1+(1+2)_decomp} is to give the basic equations by adopting the generic results of \cite{racz_geom_det} to the case of $1+(1+2)$ decompositions. The general notion of the conformal structure and the applied gauge fixing, along with the verification of the applied ideas, are given in the first part of Section \ref{spec_choice}. This section is also to provide the most important relations concerning the chosen geometrically distinguished variables. Section \ref{constraints} is to recast the constraints in terms of these variables. The analogous derivation of the evolution equations are given in Section \ref{evolutionary} but to simplify the line of argument some of the detailed derivations are moved to the Appendix. The 
mixed hyperbolic-hyperbolic system to be solved is set up in Section \ref{evolutionary_hyp_hyp} whereas the 
paper is closed by our final remarks in Section \ref{final}. 

\section{Preliminaries}\label{prelim}
\setcounter{equation}{0}

The considered four-dimensional spacetimes will be represented by a pair $(M,g_{ab})$, where $M$ is a smooth, paracompact, connected, orientable manifold $M$ endowed with a smooth Lorentzian metric $g_{ab}$.\,\footnote{All of our conventions will be as in \cite{wald}.} It is assumed that $(M,g_{ab})$ is  time orientable and that a time orientation has been chosen.

\medskip

We shall assume that Einstein's equations  hold
\begin{equation}\label{geom}
G_{ab}-\mycal{G}_{ab}=0\,,
\end{equation}
where the smooth symmetric tensor field $\mycal{G}_{ab}$---playing the role of sources---is assumed to have vanishing $\nabla^a\mycal{G}_{ab}$ divergence.

\medskip

As discussed in \cite{racz_geom_det} these assumptions host all the Einstein-matter systems with
\begin{equation}
\mycal{G}_{ab} = 8\pi\,T_{ab} - \Lambda\,g_{ab}
\end{equation}
where $T_{ab}$ denotes the energy-momentum tensor---which is known to be divergence free provided that the matter field equations hold---whereas $\Lambda$ stands for the cosmological constant. Note that (\ref{geom}) is also satisfied by the `conformally equivalent representation' of higher-curvature theories possessing a gravitational Lagrangian that is a polynomial of the Ricci scalar and it may accommodate some other alternative theories, as well.  As considerations will be restricted to the gravitational part the form of Einstein's equations as formulated by (\ref{geom}) suffices for our purposes.

\medskip

Since our spacetime $(M,g_{ab})$ is assumed to be yielded by time evolution $M$ is known to possess the product structure $M\simeq\mathbb{R}\times \Sigma$, where $\Sigma$ is some three-dimensional manifold \cite{geroch}. Then a smooth time function $\tau:M\rightarrow \mathbb{R}$ with timelike gradient such that the $\tau=const$ level surfaces $\Sigma_{\tau}=\{\tau\}\times \Sigma$ are Cauchy surfaces in $M$  is also guaranteed to exist \cite{geroch,MOSM}. Assume that a time evolution vector field $\tau^a$ has been chosen such that the relation $\tau^e\nabla_e\tau=1$ holds. 

\medskip

Loosely speaking we shall assume that the base manifold is foliated by a two-parameter family of homologous two-surfaces. What we really need to assume is that on one of the time level surfaces---say on the initial data surface $\Sigma_{0}$---there exists a smooth function $\rho: \Sigma_{0}\rightarrow \mathbb{R}$, with nowhere vanishing gradient, such that the $\rho=const$ level surfaces $\mycal{S}_\rho$ are homologous to each other. Then $\Sigma_0$ also possesses a product structure with a factor $\mycal{S}$ which can be thought of as one of the $\rho=const$ level surfaces, say $\mycal{S}=\mycal{S}_0$ . 

\medskip

The Lie transport of this foliation of $\Sigma_{0}$ along the integral curves of the time evolution vector field $\tau^a$ yields then a two-parameter foliation $\mycal{S}_{\tau,\rho}$. For simplicity, we will assume that $\Sigma_0$ and, in turn, $M$ are diffeomorphic to $\mathbb{R}\times \mycal{S}$ and $\mathbb{R}^2\times \mycal{S}$, respectively.\,\footnote{Note that more complicated topological structures could also be involved here (see, e.g.~\cite{CSPIR}). Nevertheless, as the 
geometrical issues are at the focus of our present considerations we shall restrict attention to the simplest possible topological setup.} 

\medskip

In addition to the smooth function $\rho: \Sigma_{0}\rightarrow \mathbb{R}$---by choosing suitable local coordinates\,\footnote{The spatial indices of the pull backs of geometrical objects to the $\mycal{S}_{\tau,\rho}$ surfaces will be indicated by uppercase Latin indices and they always take the values $3,4$.} $(x^3,x^4)$ on patches of $\mycal{S}_0$ and by Lie dragging the functions $x^A:\mycal{S}_{0}\rightarrow \mathbb{R}$, with $A=3,4$, along the integral curves of the vector field $\rho^a=(\partial_\rho)^a$---(local) coordinates $(\rho,x^3,x^4)$ can be defined throughout $\Sigma_{0}$.

\medskip

It is also straightforward to extend the functions $\rho, x^A$ from $\Sigma_{0}$ onto $M$ which, together with $\tau:M\rightarrow \mathbb{R}$, already fix smooth coordinates $(\tau,\rho,x^3,x^4)$ adopted to both $\tau^a$ and $\rho^a$, i.e.~$\tau^a=(\partial_\tau)^a$ and $\rho^a=(\partial_\rho)^a$. Note that while the functions $\tau$ and $\rho$ get to be defined everywhere in $M$ the spatial coordinates $x^3,x^4$ may, in general, only be defined in certain sub-domains of $M$ as $\mycal{S}_0$ may not be covered by a single map. By patching the pertinent sub-domains---based on the paracompactness of $M$, and thereby on that of $\mycal{S}_0$---all the results that can be derived, in the succeeding sections, based on coordinates defined on either of these sub-domains can always be seen to extend onto the entire of $M$. Therefore, hereafter, for the sake of simplicity, we will keep referring to the (local) coordinates $(\tau,\rho,x^3,x^4)$ as if they were defined throughout $M$.

\section{The generic $1+(1+2)$ decompositions}\label{1+(1+2)_decomp} 
\setcounter{equation}{0}

All the above assumptions concerning the topology of $M$ are to prepare the playground to perform $1+(1+2)$ decompositions which will be done in this section.

\medskip

In the conventional treatment of the Cauchy problem only a $1+3$ splitting of the metric ${g}_{ab}$ and Einstein's equations is performed. The metric ${g}_{ab}$ is decomposed as 
\begin{equation}
g_{ab}=h_{ab}- n_a n_b\,, 
\end{equation}
where $n^a$ denotes the future directed `unit norm' timelike vector field that is normal to the $\tau=const$ level surfaces, which has its own decomposition 
\begin{equation}
n^a={N}^{-1}\,\left[ (\partial_\tau)^a-N^a\right]
\end{equation}
in terms of the `laps' and `shift', $N$ and $N^a$, of the `evolution' vector field $\tau^a=(\partial_\tau)^a$ defined as 
\begin{equation}
N= -\,(\tau^e n_e) \hskip0.5cm {\rm and} \hskip0.5cm N^a={h^{a}}_{e}\,\tau^e\,,
\end{equation}
respectively.

\medskip

In spelling out all the basic relations the extrinsic curvature 
\begin{equation}\label{extcurv}
K_{ab}= {h^{e}}_{a} \nabla_e n_b=\tfrac12\,\mycal{L}_n  h_{ab}
\end{equation}
plays an important role, where $\mycal{L}_n$ stands for the Lie derivative with respect to $n^a$.

\medskip

Having all these auxiliary quantities the pertinent form of the constraint expressions and the evolution equation can be given, e.g.~by making use of (3.7), (3.8) and (3.12) of \cite{racz_geom_det} by replacing $\sigma$-derivatives there by $\tau$-derivatives and substituting $\epsilon=-1$ and $n=3$ throughout. Thus, the pull backs of the Hamiltonian and momentum constraints $E^{{}^{(\mathcal{H})}}=0$ and $E^{{}^{(\mathcal{M})}}_a=0$ to the $\Sigma_\tau$ hypersurfaces take their familiar form 
\begin{eqnarray} 
{}^{{}^{(3)}}\hskip-1mm R + \left({K^{i}}_{i}\right)^2 - K_{ij} K^{ij} - 2\,\mathfrak{e} 
{}&\hskip-.2cm=&\hskip-.2cm{} 0\,, \label{expl_eh_ch}\\
D_i {K^{i}}_{j} - D_j {K^{i}}_{i} + \mathfrak{p}_{j} {}&\hskip-.2cm=&\hskip-.2cm{} 0 \,,\label{expl_em_ch}  
\end{eqnarray}
while that of the `reduced' evolution equation (3.12) of \cite{racz_geom_det} simplifies to 
\begin{eqnarray}\label{evol_1+3} 
\hskip-1.5cm && {}^{{}^{(3)}}\hskip-1mm R_{ij}\hskip-1mm = \hskip-1mm-\mycal{L}_n K_{ij} - ({K^{l}}_{l}) K_{ij} + 2\,K_{il}{K^{l}}_{j} + {N}^{-1}\,D_i D_j N 
+[\mathfrak{S}_{ij}\hskip-1mm-\hskip-1mm\tfrac12\,h_{ij}(\mathfrak{S}_{kl}\,h^{kl}\hskip-1mm-\hskip-1mm\mathfrak{e} )] , 
\end{eqnarray}
where $D_i$ denotes the covariant derivative operator associated with $h_{ij}$, whereas $\mathfrak{e}$, $\mathfrak{p}_{a}$ and $\mathfrak{S}_{ab}$ can be given via various projections of the source term of (\ref{geom}) as 
\begin{equation}\label{1+3source}
\mathfrak{e}= n^e n^f\,\mycal{G}_{ef}\,,\hskip4mm \mathfrak{p}_{a}=-{h^{e}}_{a} n^f\, \mycal{G}_{ef}\hskip4mm {\rm and} \hskip4mm \mathfrak{S}_{ab}={h^{e}}_{a}{h^{f}}_{b}\,\mycal{G}_{ef}\,.
\end{equation}

\medskip

In economizing our assumption concerning the product structure of the spacetime manifold we may proceed as follows. Notice first that on $\Sigma_\tau$ (\ref{evol_1+3}) can be seen to be equivalent to 
\begin{equation}\label{evol_ch0}
{}^{{}^{(3)}}\hskip-1mm {G}_{ij}- {}^{{}^{(3)}}\hskip-1mm \mycal{G}_{ij}=0\,,
\end{equation}
where, in accordance with (4.3) of \cite{racz_geom_det} (with $\epsilon=-1$),
\begin{eqnarray}\label{int_evol_1+3_ch} 
&& {}^{{}^{(3)}}\hskip-1mm \mycal{G}_{ij} = \mathfrak{S}_{ij} -\mycal{L}_n K_{ij} - ({K^{l}}_{l}) K_{ij} + 2\,K_{il}{K^{l}}_{j} + {N}^{-1}\,D_i D_j N  \\
&& \phantom{ {}^{{}^{(3)}}\hskip-1mm \mycal{G}_{ij} = \mathfrak{S}_{ij}}+h_{ij}\left[\mycal{L}_n ({K^l}_{l})+\tfrac12\, ({K^{l}}_{l})^2 +\tfrac12\,K_{kl}{K^{kl}} - {N}^{-1}\,D^l D_l N \right] \nonumber \,.
\end{eqnarray}
 
Following then the generic procedure, described in Sections 2 and 3 in  \cite{racz_geom_det}, the induced metric $h_{ij}$ and 
\begin{equation}\label{evol_ch}
{}^{{}^{(3)}}\hskip-1mm E_{ij} = {}^{{}^{(3)}}\hskip-1mm {G}_{ij}- {}^{{}^{(3)}}\hskip-1mm \mycal{G}_{ij}
\end{equation}
can be decomposed---in terms of the positive definite metric $\hat \gamma_{ij}$, induced on the $\mycal{S}_{\tau,\rho}$ hypersurfaces, and the unit norm field 
\begin{equation}\label{nhat}
\hat n^i={\hat{N}}^{-1}\,[\, (\partial_\rho)^i-{\hat N}{}^i\,]
\end{equation}
normal to the $\mycal{S}_{\tau,\rho}$ hypersurfaces on $\Sigma_\tau$, where $\hat N$ and $\hat N^i$ denotes the `laps' and `shift' of the `evolution' vector field $\rho^i=(\partial_\rho)^i$ on $\Sigma_\tau$---as
\begin{equation}\label{hij}
h_{ij}=\hat \gamma_{ij}+\hat  n_i \hat n_j\,,
\end{equation}
and, by (3.7) - (3.9) of \cite{racz_geom_det}, this time with the substitution of $\epsilon=1$,  
\begin{eqnarray} 
\hat E^{{}^{(\mathcal{H})}} {}&\hskip-.4cm=&\hskip-.2cm{} \tfrac12\,\{ -\hat R + ({{\hat K}{}^{l}}{}_{l})^2 - \hat K_{kl}\hat K^{kl} - 2\,\hat{\mathfrak{e}} \}\,,\label{ham_1+(1+2)} \\
\hat E^{{}^{(\mathcal{M})}}_i {}&\hskip-.4cm=&\hskip-.2cm{} \hat D^l {{\hat K}}{}_{li} - \hat D_i {{\hat K}{}^{l}}{}_{l} - \hat{\mathfrak{p}}_{i}\,,\label{mom_1+(1+2)}\\
\hat E^{{}^{(\mathcal{EVOL})}}_{ij} {}&\hskip-.4cm=&\hskip-.2cm{} \hat R_{ij} -\mycal{L}_{\hat n} \hat K_{ij} - ({{\hat K}{}^{l}}{}_{l}) {\hat K}_{ij} + 2\,\hat K_{il}{{\hat K}{}^{l}}{}_{j} - {\hat{N}}^{-1}\,\hat D_i \hat D_j \hat N   \nonumber \\ 
&& \phantom{\hskip-.2cm{} \hat R_{ij}}
+ \hat\gamma_{ij}\{\mycal{L}_{\hat n} {\hat K}{}^l{}_{l} + \hat K_{kl}{\hat K}{}^{kl} + {\hat{N}}^{-1}\,\hat D^l \hat D_l \hat N  \} \label{evol_1+(1+2)} - [\hat{\mathfrak{S}}_{ij}-\hat{\mathfrak{e}}\,\hat \gamma_{ij}], 
\end{eqnarray}
where $\hat D_i$, $\hat R_{ij}$ and $\hat R$ denote the covariant derivative operator, the Ricci tensor and scalar curvature associated with $\hat \gamma_{ij}$, respectively.
The `hatted' source terms $\hat{\mathfrak{e}}$, $\hat{\mathfrak{p}}_{i}$ and $\hat{\mathfrak{S}}_{ij}$ and the extrinsic curvature $\hat K_{ij}$ are defined as 
\begin{equation}\label{1+3sourcehat}
\hat{\mathfrak{e}}=\hat n^k \hat n^l\,{}^{{}^{(3)}}\hskip-1mm \mycal{G}_{kl}\,, \hskip4mm \hat{\mathfrak{p}}_{i}={{\hat \gamma}^{k}}{}_{i}\,\hat n^l\, {}^{{}^{(3)}}\hskip-1mm \mycal{G}_{kl} \hskip4mm {\rm and} \hskip4mm \hat{\mathfrak{S}}_{ij}={{\hat \gamma}^{k}}{}_{i} {{\hat \gamma}^{l}}{}_{j}\,{}^{{}^{(3)}}\hskip-1mm \mycal{G}_{kl}\,,
\end{equation}
and
\begin{equation}\label{hatextcurv}
\hat K_{ij}= {{\hat \gamma}^{l}}{}_{i}\, D_l\,\hat n_j=\tfrac12\,\mycal{L}_{\hat n} {\hat \gamma}_{ij}\,.
\end{equation}

In summarizing what we have so far note that---according to the conventional $1+3$ decomposition based Cauchy problem---to get a solution to (\ref{geom}) we have to solve first the constraints (\ref{expl_eh_ch}) and (\ref{expl_em_ch}) on  the initial data surface $\Sigma_0$ and then to solve the evolutionary system (\ref{evol_ch0}) which could be replaced by the system comprised by $\hat E^{{}^{(\mathcal{H})}}=0$, $\hat E^{{}^{(\mathcal{M})}}_i=0$ and $\hat E^{{}^{(\mathcal{EVOL})}}_{ij}=0$. 

\section{The gauge choice}\label{spec_choice}
\setcounter{equation}{0}

In the previous section a combination of $1+3$ and $1+2$ decompositions was applied.  So far these decompositions were kept to be completely generic. As it was emphasized in the introduction to reduce the complexity of the Cauchy problem it is advantageous to apply a gauge fixing. This will be done in this first part of this section while in the second part the chosen geometrically distinguished variables will be characterized via certain basic relations. 

\subsection{The assumptions}\label{assumptions}  

\begin{condition}\label{cond}
In applying $1+(1+2)$ decompositions hereafter we shall assume that the followings hold:
\begin{itemize}
\item[(i)] The shift vector $N^a$ of time-evolution vector field $\tau^a=(\partial_\tau)^a$ is identically zero on $M$.
\item[(ii)] There exists a smooth non-vanishing function $\nu:M\rightarrow\mathbb{R}$ such that the lapse function of $\tau^a$ can be given by the product of $\nu$ and  the lapse function of $\rho^a$, $\hat N$, i.e. 
\begin{equation}\label{taurhoSection1}
N=\nu\,\hat N\,.
\end{equation}
\item[(iii)] There exist a smooth function $\Omega:M\rightarrow\mathbb{R}$---which does not vanish except at an origin where the foliation $\mycal{S}_{\tau,\rho}$ smoothly reduces to a point---such that the induced metric $\hat\gamma_{ij}$ can be given as 
\begin{equation}\label{thetaphiSection}
\hat\gamma_{ij} = \Omega^2\,\gamma_{ij}\,,
\end{equation}
where $\gamma_{ij}$ is such that
\begin{equation}\label{xi} 
\gamma^{ij}(\mycal{L}_\eta\gamma_{ij})=0
\end{equation}
on each of the $\mycal{S}_{\tau,\rho}$ surfaces, where $\eta^a$ stands for either of the coordinate basis fields  $\tau^a=(\partial_\tau)^a$ or $\rho^a=(\partial_\rho)^a$.
\end{itemize}
\end{condition}

Some comments are in order now to verify the consistency of the above conditions. 
Note first that the requirement specified in $(i)$ guaranties that $\tau^a$ is parallel to the unit normal, i.e.~it is also normal to the $\Sigma_\tau$ hypersurfaces and the relation
\begin{equation}\label{(i)} 
n^a={N}^{-1}\,\tau^a
\end{equation}
holds. 

\medskip

For the first glance this assumption may look to be too strong. Nevertheless, the spacetime metric possessing this form {is sufficiently generic and it} is applied in various investigations.  For instance in \cite{chris} it is referred as the canonical form of the spacetime metric. More importantly, as it was justified recently by M\"uller and S\'anches \cite{MOSM} the gauge choice with vanishing shift may indeed be considered as a {generic} one. More precisely, it was shown in \cite{MOSM} that to any globally hyperbolic spacetime $(M,g_{ab})$ there always exists a smooth time function $\tau:M\rightarrow \mathbb{R}$ with timelike gradient such that the $\tau=const$ level surfaces are Cauchy surfaces, and also the metric can be given in the form 
\begin{equation}\label{pref}
g_{ab}=-N^2\, (d\tau)_a(d\tau)_b+h_{ab}\,,
\end{equation} 
with a bounded lapse function $N:M\rightarrow \mathbb{R}$, and with a smooth Riemannian metric $h_{ab}$ on the $\Sigma_{\tau}$ time level surfaces. 

\medskip

Once an attempt is made to solve an initial value problem it is crucial to find conditions guaranteeing the maximality of the pertinent Cauchy development. For a related discussion and for some explicit examples see pages 5-6 of \cite{CSPIR} where it is demonstrated that a bad choice of lapse---regardless what sort of shift is used---could get on the way of the acquiring the maximal development. The above recalled result of M\"uller and S\'anches \cite{MOSM} is of fundamental importance as it guaranties that within the set of spacetimes with canonical metric all the possible globally hyperbolic spacetimes can appropriately be represented. In particular, each of the maximal Cauchy developments has a gauge equivalent representation within this set{, and, as it will be shown in Section \ref{constraints}, the pertinent constraint equations can be solved in the generic case in a very effective way.}

\medskip

Note that condition $(ii)$ automatically holds as both of the lapses $N$ and $\hat N$ are smooth and non-vanishing by construction so the existence of $\nu$ is guaranteed. This condition will be applied in spelling out the explicit form of our evolutionary equations in Section \ref{evolutionary}. Note also that conditions $(i)$ and $(ii)$ are consistent with the general expectation that the dynamical fields are represented by the components of the three-metric on the $\Sigma_{\tau}$ time level surfaces \cite{MOSM}.

\medskip

In short terms condition $(iii)$ is also to select our geometrically distinguished variables. In the terminology proposed by Stachel and d'Inverno \cite{diverno1} the metric $\gamma_{ij}$ {in} (\ref{thetaphiSection}) is supposed to represent the `conformal structure'. Note, however, that in spite of the central role of this notion they did not attempt to give a clear geometric determination of $\Omega$ and $\gamma_{ij}$ in a spacetime sense. 

\medskip

In providing a constructional definition of both of these objects we may proceed as follows.
Recall first that the $\mycal{S}_{\tau,\rho}$ surfaces are homologous to each other. To show that the desired smooth function $\Omega:M\rightarrow\mathbb{R}$ and the metric $\gamma_{ij}$ exist one may apply the identity
\begin{equation}\label{xi_2}
\hat\gamma^{ij}(\mycal{L}_\eta\hat\gamma_{ij})=\gamma^{ij}(\mycal{L}_\eta\gamma_{ij})+2\,\mycal{L}_\eta(\ln \Omega^2)\,,
\end{equation}
where $\eta^a$ stands either for $\tau^a$ or for $\rho^a$. 

Assuming now that the first term on the right hand side of (\ref{xi_2}) vanishes\,\footnote{Notice that then 
\begin{equation}\label{xi_2_2}
\hat\gamma^{ij}(\mycal{L}_\eta\hat\gamma_{ij})=2\,\mycal{L}_\eta(\ln \Omega^2)=2\,\hat\theta_{(\eta)}\,,
\end{equation}
where $\hat\theta_{(\eta)}$ measures the rate of change of the two-volume, associated with $\hat\gamma_{ij}$, in the direction of $\eta^a$.} for any smooth distribution of the induced two-metric $\hat\gamma_{ij}$ on the $\mycal{S}_{\tau,\rho}$ surfaces one may integrate (\ref{xi_2}) first along the integral curves of $\rho^a$ on $\Sigma_0$, starting at $\mycal{S}_{0}$, and then along the integral curves of $\tau^a$, starting at the surface $\mycal{S}_{\rho}$ {on} $\Sigma_0$. The functional form of $\Omega^2=\Omega^2(\tau,\rho,x^3,x^4)$ reads as 
\begin{equation}\label{xi_om}
\Omega^2=\Omega^2_0\cdot\exp\left[\tfrac12\int_0^\rho \left(\hat\gamma^{ij}(\mycal{L}_\rho\hat\gamma_{ij})\right)\,d\tilde\rho \right]\cdot\exp\left[\tfrac12\int_0^\tau \left(\hat\gamma^{ij}(\mycal{L}_\tau\hat\gamma_{ij})\right)\,d\tilde\tau \right]\,,
\end{equation}
where $\Omega_0=\Omega_0(x^3,x^4)$ denotes the conformal factor at $\mycal{S}_{0}$.\,\footnote{Note that to have $\Omega$ being well-defined throughout $M$ the above integrations are meant to be done in both directions along the integral curves of $\rho^a$ and $\tau^a$, respectively.}

\medskip

Unless $\mycal{S}_{0}$ represents an origin $\Omega_0$ does not vanish on $\mycal{S}_{0}$ neither does $\Omega$ in a sufficiently small neighborhood of $\mycal{S}_{0}$. Note, however, that the limiting behavior $\hat\gamma^{ij}(\mycal{L}_\rho\hat\gamma_{ij})\rightarrow \pm\infty$ of the integrand of (\ref{xi_om}) while  $\rho\rightarrow\rho_*^\pm$ is a clear indication of the existence of an origin located at $\rho=\rho_*$ characterized also by the vanishing of the limit of  $\Omega$ there.

\medskip

It is also important to know whether $\Omega$, yielded by the above process, is indeed consistently defined throughout $M$. In other words, it has to be checked whether we would ended up with the same conformal factor on the individual surfaces $\mycal{S}_{\tau,\rho}$ if we started the  integration of (\ref{xi_2}) first along the integral curves of $\tau^a$ set out at the points of $\mycal{S}_{0}$ and then integrated along the integral curves of $\rho^a$ on $\Sigma_\tau$. To see that $\Omega:M\rightarrow\mathbb{R}$ as given by  (\ref{xi_om}) is a well-defined smooth function we may refer to the integrability condition for $\ln \Omega^2$,
\begin{equation}\label{xi_3}
\mycal{L}_\tau\left[\hat\gamma^{ij}(\mycal{L}_\rho\hat\gamma_{ij})\right]-\mycal{L}_\rho\left[\hat\gamma^{ij}(\mycal{L}_\tau\hat\gamma_{ij})\right]=0\,,
\end{equation}
which holds as the vector fields $\tau^a$ and $\rho^a$ do commute by construction.

\medskip

Once we have the smooth function $\Omega:M\rightarrow\mathbb{R}$ the smooth distribution of two-metrics $\gamma_{ij}$ satisfying all the requirements posed in Condition \ref{cond} can be determined---except at an origin where one should take the limit, $\lim_{\rho\rightarrow\rho_*}\gamma_{ij}$---, in virtue of (\ref{thetaphiSection}), in terms of smoothly varying induced metric $\hat\gamma_{ij}$ and $\Omega$ on $\mycal{S}_{\tau,\rho}$. 

\medskip

Recall also that once $\Omega$ and $\gamma_{ij}$ are defined on the $\mycal{S}_{\tau,\rho}$ level surfaces the scalar curvature $\hat R$ can be given as 
\begin{equation}\label{hatextcurv2}
\hat R = \Omega^{-2}\left[{}^{{}^{(\gamma)}}\hskip-1mm R-\mathbb{D}^l\mathbb{D}_l \ln\Omega^2\right]\,,
\end{equation}
where ${}^{{}^{(\gamma)}}\hskip-1mm R$ and $\mathbb{D}_i$ denote the scalar curvature and the covariant derivative operator associated with $\gamma_{ij}$, respectively.\,\footnote{The indices in expressions involving the covariant derivative operator $\mathbb{D}_i$ will always be raised and lowered by $\gamma^{ij}$ and $\gamma_{ij}$, respectively. Note also that elsewhere the indices are raised and lowered by $\hat\gamma^{ij}$ and $\hat\gamma_{ij}$, respectively}. Note also that $\mathbb{D}_i$ and $\hat D_i$ are related via the $(1,2)$ type tensor 
\begin{equation}
{C^k}_{ij}= {\delta^k}_{(i}\mathbb{D}_{j)}\ln\Omega^2-\tfrac12\,\gamma_{ij}\,\mathbb{D}^k\ln\Omega^2\,.
\end{equation}
This means, for instance, that the $\hat D_i$-derivative of a one-form field $Z_i$ on the $\mycal{S}_{\tau,\rho}$ surfaces is given as
\begin{equation}
\hat D_i\,Z_j= \mathbb{D}_i\,Z_j - {C^k}_{ij}\,Z_k\,.
\end{equation}
It is worth keeping in mind that once ${\gamma}_{ij}$ is known its scalar curvature ${}^{{}^{(\gamma)}}\hskip-1mm R$ is also a known function on each of the $\mycal{S}_{\tau,\rho}$ level surfaces. 

\medskip

In summarizing note that in virtue of Condition \ref{cond} our dependent variables can be listed as $\Omega, \gamma_{ij}, \hat N, \hat N^i;\nu$.  As we will see no evolution equation applies to $\nu$ thereby only the first six of these variables---determining the three-metric $h_{ij}$---will play dynamical role while $\nu$ will turn out to play the role of a `gauge source function'.

\subsection{The basic variables}\label{useful_relations} 

As a preparation for the derivation of the field equations it is rewarding to have a closer view of our basic variables. In doing so recall first that, in virtue of (\ref{hij}), the three-metric $h_{ij}$ in adopted (local) coordinates $(\tau,\rho,x^3,x^4)$ reads as 
\begin{equation}\label{hijind}
h_{ij}= (\hat N^2+ \hat N_E\hat N^E)\,({\rm d}\rho)_i({\rm d}\rho)_j+2\,\hat N_A\,({\rm d}\rho)_{(i}({\rm d} x^A)_{j)} +  \hat\gamma_{AB}\,({\rm d} x^A)_i\,({\rm d} x^B)_j\,.
\end{equation}
The extrinsic curvature $K_{ij}=\tfrac12\,{\mycal{L}_n} h_{ij}$ can then be given as 
\begin{equation}\label{Kij_ind}
2\,K_{ij}= \mycal{L}_n (\hat N^2+ \hat N_E\hat N^E)\, ({\rm d}\rho)_i({\rm d}\rho)_j+2\,\mycal{L}_n \hat N_A\,({\rm d}\rho)_{(i}({\rm d} x^A)_{j)} +  \mycal{L}_n \hat\gamma_{AB}\,({\rm d} x^A)_i\,({\rm d} x^B)_j\,
\end{equation}
since, in virtue of (\ref{(i)}), in the adopted coordinates $(\tau,\rho,x^3,x^4)$ for the Lie derivative of coordinate differentials with respect to $n^a$
\begin{equation}
\mycal{L}_n ({\rm d} x^\alpha)_i ={N}^{-1} \,\mycal{L}_\tau ({\rm d} x^\alpha)_i =0\,
\end{equation}
hold. 

\medskip

By applying then the relations 
\begin{equation}
\hat n_i= \hat N ({\rm d}\rho)_i\ \ \ {\rm and}\ \ \ \hat \gamma^i{}_j=\delta^i{}_j-\hat n^i\hat n_j
\end{equation}
we get
\begin{equation}
\hat n^i({\rm d}\rho)_i= \hat N^{-1}\,,\ \ \ \hat n^i({\rm d} x^A)_i= -\hat N^{-1}\hat N^A\ \ \ {\rm and}\ \ \ ({\rm d}\rho)_j\hat \gamma^j{}_i=0\,,
\end{equation}
which, along with (\ref{Kij_ind}) and the decomposition
\begin{equation}\label{dec_1}
K_{ij}= \boldsymbol\kappa \,\hat n_i \hat n_j  + \left[\hat n_i \,{\rm\bf k}{}_j  + \hat n_j\,{\rm\bf k}{}_i\right]  + {\rm\bf K}_{ij}\,,
\end{equation}
of the extrinsic curvature imply that for the boldface quantities the relations
\begin{eqnarray}
\hskip-.2cm\boldsymbol\kappa {}&\hskip-.2cm=&\hskip-.2cm{} \hat n^k\hat  n^l\,K_{kl} = \mycal{L}_n \ln \hat N \label{loc_expr41} \\
\hskip-.2cm{\rm\bf k}{}_{i} {}&\hskip-.2cm=&\hskip-.2cm{} {\hat \gamma}^{k}{}_{i}\hat  n^l\, K_{kl} = {(2\hat N)}^{-1}\,\hat \gamma_{AB}\,(\mycal{L}_{n}\hat N^B)\,[({\rm d} x^A)_k\,\hat \gamma^k{}_i] = {(2\hat N)}^{-1}\,\hat \gamma_{il}\,(\mycal{L}_{n}\hat N^l)  \label{loc_expr42}\\
\hskip-.2cm{\rm\bf K}_{ij} {}&\hskip-.2cm=&\hskip-.2cm{} {\hat \gamma}^{k}{}_{i} {\hat \gamma}^{l}{}_{j}\,K_{kl} = \tfrac12\,\mycal{L}_{n}\hat \gamma_{AB}\,[({\rm d} x^A)_k\,\hat \gamma^k{}_i] \,[({\rm d} x^A)_l\,\hat \gamma^l{}_j] = \tfrac12\,\hat \gamma^k{}_i\hat \gamma^l{}_j\,(\mycal{L}_{n}\hat \gamma_{kl})\, \label{loc_expr43}
\end{eqnarray}
hold. 
Notice that all these boldface quantities may be considered as if they were defined exclusively on the $\mycal{S}_{\tau,\rho}$ level surfaces. 
Note also that $[({\rm d} x^A)_k\,\hat \gamma^k{}_i]=({\rm d} x^A)_i+\hat N^A\,({\rm d}\rho)_i$ holds, whence the pull-back of $[({\rm d} x^A)_k\,\hat \gamma^k{}_i]$ to the $\mycal{S}_{\tau,\rho}$ surfaces, in particular, when it is written in coordinates $(\tau,\rho,x^3,x^4)$ adopted to the foliation $\mycal{S}_{\tau,\rho}$ and the vector fields $\tau^a$ and $\rho^a$, simply takes the form $\delta^A{}_i$. 

\medskip

Based on the above introduced decomposition we also have the relations 
\begin{equation}\label{dec_2}
{K^l}{}_{l}=\boldsymbol\kappa+{\rm\bf K}^l{}_{l}\,,\ \ \ {\rm and }\ \ \ K_{kl}K^{kl}= \boldsymbol\kappa^2+2\,{\rm\bf k}^{l}{\rm\bf k}{}_{l}+ {\rm\bf K}_{kl}{\rm\bf K}^{kl}\,,
\end{equation}
where the indices of the boldface quantities are (effectively) raised and lowered by $\hat \gamma^{ij}$ and $\hat \gamma_{ij}$, respectively.  

\medskip

Some of the above contractions can also be given in terms of various derivatives of the geometrically distinguished variables $\Omega, \gamma_{ij}, \hat N, \hat N^i$. For instance, in virtue of (\ref{thetaphiSection}), (\ref{xi}), (\ref{xi_2}), (\ref{loc_expr41}) and (\ref{dec_2}), the trace $K^l{}_l=h^{kl}K_{kl}$ can be given as
\begin{equation}\label{loc_expr8}
K^l{}_l= \mycal{L}_n \ln \hat N + \mycal{L}_n \ln \Omega^2\,.
\end{equation}

\medskip

Analogously, in virtue of (\ref{hatextcurv}), we have that
\begin{equation}\label{hatK}
\hat K_{ij}= \tfrac12\left[ (\mycal{L}_{\hat n} \ln  \Omega^2)\,\hat \gamma_{ij} + \Omega^2\, \mycal{L}_{\hat n}\gamma_{ij}\right] \,,
\end{equation}
from which for the trace $\hat K^l{}_l=\hat K_{kl}\hat \gamma^{kl}$ we get
\begin{equation}\label{trhatK}
\hat K{}^l{}_l= \mycal{L}_{\hat n} \ln \Omega^2+\tfrac12\,\gamma^{kl}\mycal{L}_{\hat n}\gamma_{kl}=\mycal{L}_{\hat n} \ln \Omega^2-{\hat{N}}^{-1}\,\mathbb{D}_k \hat N^k\,,
\end{equation}
where in the last step (\ref{nhat}) and  (\ref{xi}), with $\eta^a=\rho^a$, were applied. Note that $\hat K^l{}_l$ may also be written as 
\begin{equation}\label{trhatK2}
\hat K{}^l{}_l= {\hat{N}}^{-1}\,[\mycal{L}_{\rho} \ln \Omega^2-\hat {D}_k \hat N^k]\,.
\end{equation}

\medskip

It is also advantageous to define the (conformal invariant) operator 
\begin{equation}\label{PI_def}
\Pi^{kl}{}_{ij}=\hat\gamma^{k}{}_{i}\hat\gamma^{l}{}_{j}-\tfrac12\,\hat\gamma_{ij}\hat\gamma^{kl}=\gamma^{k}{}_{i}\gamma^{l}{}_{j}-\tfrac12\,\gamma_{ij}\gamma^{kl}\,
\end{equation}
mapping symmetric tensors defined on $\Sigma_\tau$ to their traceless symmetric part such that the resulted fields may again be viewed as if they were fields on the surfaces $\mycal{S}_{\tau,\rho}$, exclusively.\,\footnote{ It is straightforward to {verify} that 
\begin{equation}\label{PI_negyzet}
\Pi^{kl}{}_{ij}\Pi^{pq}{}_{kl}=\Pi^{pq}{}_{ij}\,,
\end{equation}
and that for an arbitrary symmetric tensor field $\mycal{A}_{kl}$ on $\mycal{S}_{\tau,\rho}$ that contains a part proportional to $\hat \gamma^{kl}$, i.e.~reads as  
\begin{equation}
\mycal{A}_{kl}=[\mycal{A}]_{kl}+\hat \gamma_{kl}\,\{\mycal{A}\}\,,
\end{equation}
the relation 
\begin{equation}\label{PI2}
\Pi^{kl}{}_{ij}\,\mycal{A}_{kl}=[\mycal{A}]_{ij}-\tfrac12\,\hat \gamma_{ij}\,(\hat \gamma^{kl}\,[\mycal{A}]_{kl})\,,
\end{equation}
holds. Thus in evaluating $\Pi^{kl}{}_{ij}\,\mycal{A}_{kl}$ the part of $\mycal{A}_{kl}$ explicitly proportional to $\hat \gamma_{kl}$ can always be left out of considerations.}
In particular, it can be verified that the trace-free projection of the extrinsic curvatures  
\begin{equation}
\Pi^{kl}{}_{ij}\,K_{kl}=[\hat\gamma^{k}{}_{i}\hat\gamma^{l}{}_{j}-\tfrac12\,\hat\gamma_{ij}\hat\gamma^{kl}]\,K_{kl}  \ \ \ {\rm and }\ \ \ 
\Pi^{kl}{}_{ij}\,\hat K_{kl}=[\hat\gamma^{k}{}_{i}\hat\gamma^{l}{}_{j}-\tfrac12\,\hat\gamma_{ij}\hat\gamma^{kl}]\,\hat K_{kl}
\end{equation}
read, by making use of the handy notation $\Pi^{kl}{}_{ij}\,K_{kl}=\interior{\rm\bf K}_{ij}$ and $\Pi^{kl}{}_{ij}\,\hat K_{kl}=\interior{\hat K}{}_{ij}$,  as 
\begin{equation}\label{interiors0}
\interior{\rm\bf K}_{ij}={\rm\bf K}_{ij}-\tfrac12\,\gamma_{ij}(\gamma^{ef}{{\rm\bf K}}_{ef}) \ \ \ {\rm and }\ \ \ 
\interior{\hat K}{}_{ij}={\hat K}_{ij}-\tfrac12\,\gamma_{ij}(\gamma^{ef}{{\hat K}}_{ef})\,,
\end{equation}
which can be related to the Lie derivatives of the conformal structure $\gamma_{ij}$ as
\begin{equation}\label{interiors}
\interior{\rm\bf K}_{ij}=\tfrac12\,{\Omega^2}\,\hat \gamma^k{}_i\hat \gamma^l{}_j\,(\mycal{L}_n \gamma_{kl}) \ \ \ {\rm and }\ \ \ 
\interior{\hat K}_{ij}=\tfrac12\,{\Omega^2}\,[\mycal{L}_{\hat n} \gamma_{ij}-{\hat N}^{-1}\,\gamma_{ij} (\mathbb{D}_l \hat N^l)]\,.
\end{equation}
Note that, in virtue of (\ref{interiors0}), the relations
\begin{equation}\label{dec_4}
{\rm\bf K}_{ij}{\rm\bf K}^{ij}=\tfrac12\,({\rm\bf K}^l{}_{l})^2 + \interior{\rm\bf K}{}_{ij}\,\interior{\rm\bf K}{}^{ij}  \ \ \ {\rm and }\ \ \
{\hat K}_{ij}{\hat K}^{ij}=\tfrac12\,({\hat K{}^l}{}_{l})^2 + \interior{\hat K}{}_{ij}\,\interior{\hat K}{}^{ij} \,,
\end{equation}
along with 
\begin{equation}\label{dec_5}
\hat D^i{\rm\bf K}_{ij}=\tfrac12\, \hat D_j({\rm\bf K}^l{}_{l})+ \hat D^i\,\interior{\rm\bf K}{}_{ij}\,,
\end{equation}
are satisfied.

\medskip

In closing this section let us mention that $\mycal{L}_n K_{ij}$ can also be expressed by an equation analogous to (\ref{Kij_ind}) with the distinction that the first order Lie derivatives with respect to $n^a$ have to be replaced by second order ones. By exactly the same type of argument as applied in deriving (\ref{hijind})-(\ref{loc_expr43}) the relations 
\begin{eqnarray}
\hskip-9mm \hat n^k\hat  n^l(\mycal{L}_n K_{kl}) {}&\hskip-.25cm=&\hskip-.25cm{} \mycal{L}^2_n \ln \hat N+2\,[\mycal{L}_n \ln \hat N]^2 + {\hat N{}^{-2}}\,\hat \gamma_{kl}  (\mycal{L}_{n}\hat N^k)(\mycal{L}_{n}\hat N^l)\,,\label{loc_expr_liek1}\\
\hskip-9mm {\hat \gamma}^{k}{}_{i}\hat  n^l (\mycal{L}_n K_{kl}) {}&\hskip-.25cm=&\hskip-.25cm{} {\hat{N}}^{-1}\left[\tfrac12\,\hat \gamma_{il}(\mycal{L}^2_{n}\hat N^l)+{\Omega^2}(\mycal{L}_{n} \gamma_{il})(\mycal{L}_{n}\hat N^l)+ \hat\gamma_{il}(\mycal{L}_{n}\hat N^l)(\mycal{L}_n \ln \Omega^2) \phantom{\tfrac12}\hskip-.3cm\right] \label{loc_expr_liek2} \\
\hskip-9mm {\hat \gamma}^{k}{}_{i} {\hat \gamma}^{l}{}_{j}(\mycal{L}_n K_{kl}) {}&\hskip-.25cm=&\hskip-.25cm{} \tfrac12\,\hat \gamma^k{}_i\hat \gamma^l{}_j\,(\mycal{L}^2_{n}\hat \gamma_{kl})\,.\label{loc_expr_liek3}
\end{eqnarray}
can also be seen to hold.

\section{The $1+3$ constraints}\label{constraints} 
\setcounter{equation}{0}

This section is to recast the $1+3$ Hamiltonian and momentum constraints,  given by (\ref{expl_eh_ch}) and (\ref{expl_em_ch}), into a coupled parabolic-hyperbolic system on $\Sigma_0$ to which---whenever the foliating two-surfaces are compact without boundary in the spacetime---an initial value problem applies.

\subsection{The momentum constraint}\label{constraints1st}

Taking the parallel and orthogonal projections of (\ref{expl_em_ch}), respectively, and a straightforward adaptation of (A.9), (A.10), (A.12) and (A.13) in the Appendix of \cite{racz_geom_det}, along with (\ref{dec_2}) - (\ref{dec_5}), we get 
\begin{eqnarray}
({\hat K^{l}}{}_{l})\,{\rm\bf k}{}_{i} + \hat D^l \interior{\rm\bf K}{}_{li} + \boldsymbol\kappa\,\dot{\hat n}{}_i +\mycal{L}_{\hat n} {\rm\bf k}{}_{i} - \dot{\hat n}{}^l\,{\rm\bf K}_{li} 
- \hat D_i\boldsymbol\kappa - \tfrac12\,\hat D_i ({\rm\bf K}^l{}_{l}) + \mathfrak{p}_l\,{\hat \gamma^{l}}{}_{i} {}&\hskip-.4cm=&\hskip-.2cm{} 0 \label{par_const} \\
\boldsymbol\kappa\,({\hat K^{l}}{}_{l}) + \hat D^l {\rm\bf k}_{l}-\tfrac12\,({\rm\bf K}^l{}_{l})\,({\hat K^{k}}{}_{k}) - \interior{\rm\bf K}{}_{kl}\interior{\hat K}{}^{kl}  -2\,\dot{\hat n}{}^l\, {\rm\bf k}_{l}  - \mycal{L}_{\hat n}({\rm\bf K}^l{}_{l})+\mathfrak{p}_l\,{\hat n^{l}} {}&\hskip-.4cm=&\hskip-.2cm{} 0\,, \label{ort_const}
\end{eqnarray}
{where $\dot{\hat n}{}_k={\hat n}{}^lD_l{\hat n}{}_k=-{\hat D}_k(\ln{\hat N})$.}

Taking then derivatives of (\ref{loc_expr42}) we get 
\begin{align}
\mycal{L}_{\hat{n}}\mathbf{k}_{i}={}&{(2 \hat{N})}^{-1}{\hat{\gamma}_{ij}\,[\mathcal{L}_{\hat{n}}(\mathcal{L}_{n}\hat{N}^{j})]} + 2\,\hat{K}_{ij}\,\mathbf{k}^{j}- \mycal{L}_{\hat{n}}(\ln\hat{N})\,\mathbf{k}_{i} \\
\hat{D}^{l}\mathbf{k}_{l}={}&{(2 \hat{N})}^{-1}{\hat{D}_{l}(\mathcal{L}_{n}\hat{N}^{l})}
 - \mathbf{k}^{l}\,\hat{D}_{l}(\ln\hat{N})\,, 
\end{align}
whence (\ref{par_const}) and (\ref{ort_const}) can be recast as
\begin{align}
\hskip-0.5cm({\hat K^{l}}{}_{l})\,{\rm\bf k}{}_{i} + \hat D^l \interior{\rm\bf K}{}_{li} + \boldsymbol\kappa\,\dot{\hat n}{}_i +{(2 \hat{N})}^{-1}{\hat{\gamma}_{ij}\,[\mathcal{L}_{\hat{n}}(\mathcal{L}_{n}\hat{N}^{j})]} + 2\,\hat{K}_{ij}\mathbf{k}^{j}- \mycal{L}_{\hat{n}}(\ln\hat{N}) \mathbf{k}_{i}   & \nonumber\\
- \dot{\hat n}{}^l\,{\rm\bf K}_{li} - \hat D_i\boldsymbol\kappa - \tfrac12\,\hat D_i (\mathcal{L}_{n}\ln\Omega^2) + \mathfrak{p}_l\,{\hat \gamma^{l}}{}_{i}{}&={} 0  \label{par_const2} \\
\boldsymbol{\kappa}\,\hat{K}^{l}{}_{l} -   \mathbf{K}_{kl}\hat{K}^{kl}
 + {(2 \hat{N})}^{-1}{\hat{D}_{l}(\mathcal{L}_{n}\hat{N}^{l})}
 -  \mathcal{L}_{\hat{n}} [\mathcal{L}_{{n}}(\ln\Omega^2)]+ \mathbf{k}^{l}\,\hat{D}_{l}(\ln\hat{N}) + \mathfrak{p}_{l}\hat{n}^{l} 
{}&={} 0.\label{ort_const2}
\end{align}

Remarkably, twice of (\ref{par_const2}) and $-2\,\hat N$ times of (\ref{ort_const2})---when writing them out in coordinates $(\rho,x^3,x^4)$ adopted to the foliation $\mycal{S}_{\rho}$ and the vector field $\rho^i$ on $\Sigma_0$---can be seen to take the form 
\begin{equation}\label{constr_hyp}
\left\{ \left(\hskip-.09cm
\begin{array}{cc}
 \hskip-.09cm \frac{1}{\hat N{}^2}{\hat \gamma}_{AB} & \hskip-.09cm 0 \\ 
 \hskip-.09cm 0  &\hskip-.09cm 2 
\end{array} 
\hskip-.09cm \right)\,\partial_\rho +
\left(\hskip-.09cm
\begin{array}{cc}
 \hskip-.09cm -  \frac{\hat N^{K}}{\hat N{}^2}{\hat \gamma}_{AB} & \hskip-.09cm- {\hat \gamma}_A{}^{K}  \\
 \hskip-.09cm - {\hat \gamma}_B{}^{K} &  \hskip-.09cm -  2\,\hat N^{K}
\end{array} \hskip-.09cm
\right)\,\partial_K
\right\}
\left(\hskip-.09cm
\begin{array}{c}
\mathcal{L}_{n}\hat{N}^{B} \\
\mathcal{L}_{n} \ln\Omega^2
\end{array} \hskip-.09cm
\right) +
\left(\hskip-.09cm
\begin{array}{c}
{}^{{}^{(\mathcal{L}_{n}\hat{N}^{B})}}\hskip-1mm\mycal{B}_{\,A} \\
{}^{{}^{(\mathcal{L}_{n} \ln\Omega^2)}}\hskip-1mm\mycal{B}
\end{array} \hskip-.09cm
\right)=0\,.
\end{equation}
Since the coefficients of the derivatives $\partial_\rho$ and $\partial_K$ are symmetric and the coefficient of $\partial_\rho$ is also positive definite (\ref{constr_hyp}) can be seen to form a first order symmetric hyperbolic system 
\begin{equation}\label{constr_hyp2}
\mathcal A^{(\rho)} \,\partial_\rho {\bf u} + \mathcal A^{(K)} \,\partial_K {\bf u}+ \mathcal B = 0 
\end{equation}
(with $K=3,4$) for the vector valued variable 
\begin{equation}\label{vector-valued}
{\bf u}=(\mathcal{L}_{n}\hat{N}^{B}, \mathcal{L}_{n} \ln\Omega^2)^T\,,
\end{equation}
where the `radial coordinate' $\rho$ play{s} now the role of time.\,\footnote{It is worth emphasizing that, without referring to the metric tensor components, (\ref{par_const}) and (\ref{ort_const}) as they stand---when writing them out in coordinates $(\rho,x^3,x^4)$ adopted to the foliation $\mycal{S}_{\rho}$ and the vector field $\rho^i$ on $\Sigma_0$---can be put into a first order symmetric hyperbolic system for the vector valued variable $({\rm\bf k}{}_{A}, {\rm\bf K}^E{}_{E})^T$, involving the projections of the extrinsic curvature $K_{ij}$ exclusively.} 

\medskip

Notice that in virtue of (\ref{par_const2}) and (\ref{ort_const2}) the coefficients of the principal part and the components ${}^{{}^{(\mathcal{L}_{n}\hat{N}^{B})}}\hskip-1mm\mycal{B}_{\,A}$ and ${}^{{}^{(\mathcal{L}_{n} \ln\Omega^2)}}\hskip-1mm\mycal{B}$ of the source $\mathcal B$ are comprised by smooth terms which individually may be considered as fields defined on the level surfaces $\mycal{S}_{\rho}$. 

\medskip

It is also informative to inspect the characteristic polynomial which reads as 
\begin{equation}
\sigma(x^i,\xi^i)=2\,\det(h_{ij})\,(\hat n^i \xi_i)\,(\hat \gamma{}^{ij} - 2\,\hat n^i \hat n^j)\,\xi_i\xi_j \,,
\end{equation}
i.e.~the `characteristic cone' associated with (\ref{constr_hyp})---apart from the hypersurfaces $\Sigma_\rho$ with $\hat n^i \xi_i = 0$---is given as
\begin{equation}
(\hat \gamma{}^{ij} - 2\,\hat n^i \hat n^j)\, \xi_i \xi_j =  0\,.
\end{equation}
Accordingly, we have 
\begin{proposition}\label{mom_contsr}
The momentum constraint can be given by (\ref{par_const2}) and (\ref{ort_const2}). When writing these equations out in coordinates $(\rho,x^3,x^4)$ adopted to the foliation $\mycal{S}_{\rho}$ and the vector field $\rho^i$ on $\Sigma_0$ they form a first order symmetric hyperbolic system (\ref{constr_hyp}) for the vector valued variable $\bf u$ as given in (\ref{vector-valued}). The coefficients $\mathcal A^{(\rho)}$ and $\mathcal A^{(K)}$ of  (\ref{constr_hyp}), along with its source term $\mathcal B$, smoothly depend on $\bf u$ and quantities defined on the $\mycal{S}_{\rho}$ hypersurfaces in $\Sigma_0$.  
\end{proposition}  

\subsection{The $1+3$ Hamiltonian constraint}\label{constraints2nd} 

This subsection is to show that the Hamiltonian constraint on $\Sigma_0$ can be put into the form of a second order elliptic system for the conformal factor $\Omega$ or, alternatively, into the form of a parabolic equations for the lapse function $\hat N$.

\medskip

Notice first that by (A.1) of \cite{racz_geom_det} 
\begin{equation}
{}^{{}^{(3)}}\hskip-1mm R= \hat R - 2\,\mycal{L}_{\hat n} ({\hat K^l}{}_{l})-({\hat K^{l}}{}_{l})^2 - \hat K_{kl} \hat K^{kl} - 2\,{\hat N}^{-1}\,\hat D^l \hat D_l \hat N 
\end{equation}
holds. By combining this with (\ref{dec_2})-(\ref{dec_4}) {twice of} the Hamiltonian constraint (\ref{expl_eh_ch}) can be written as
\begin{eqnarray}
&& \hskip-1.cm \hat R - 2\,\mycal{L}_{\hat n} ({\hat K^l}{}_{l})-\tfrac32\,({\hat K^{l}}{}_{l})^2 - \interior{\hat K}{}_{kl}\,\interior{\hat K}{}^{kl} - 2\,{\hat N}^{-1}\,\hat D^l \hat D_l \hat N +2\,\boldsymbol\kappa\,({\rm\bf K}^l{}_{l})+\tfrac12({\rm\bf K}^l{}_{l})^2 \nonumber \\ && \phantom{\  - 2\,\mycal{L}_{\hat n} ({\hat K^l}{}_{l})- \interior{\hat K}{}_{kl}\,\interior{\hat K}{}^{kl} - 2\,{\hat N}^{-1}\,\hat D^l \hat D_l \hat N } 
-2\,{\rm\bf k}{}^{l}{\rm\bf k}{}_{l}- \interior{\rm\bf K}{}_{kl}\,\interior{\rm\bf K}{}^{kl} -2\,\mathfrak{e}=0\,. \label{scal_const}
\end{eqnarray}

It may be tempting to consider (\ref{scal_const}) also as a first order partial differential equation for the variable ${\hat K^{l}}{}_{l}$ or, more appropriately, for $\mycal{L}_{\hat n}\ln\Omega^2$. Note, however, that because of the presence of the scalar curvature $\hat R$ in (\ref{scal_const}), especially in virtue of (\ref{hatextcurv2}), there are hidden higher order derivatives of $\Omega$ in (\ref{scal_const}).

\medskip

Using (\ref{hatK}), (\ref{trhatK}) and (\ref{hatextcurv2}) one can uncover these derivatives and, in turn, the Hamiltonian constraint can be seen to take the form 
\begin{eqnarray}\label{scal_constr_ell}
&& \hskip-1.7cm \Omega^{-2}\left[{}^{{}^{(\gamma)}}\hskip-1mm R - \mathbb{D}^k\mathbb{D}_k \ln \Omega^2\right] - 2\,\mycal{L}_{\hat n}\left[ \mycal{L}_{\hat n}\,\ln \Omega^2 -{\hat{N}}^{-1}\,(\mathbb{D}_k \hat N{}^k)\right]-\tfrac32\,({\hat K^{l}}{}_{l})^2 - \interior{\hat K}{}_{kl}\,\interior{\hat K}{}^{kl} \nonumber \\ && \hskip-1.0cm \phantom{\Omega^{-2}\,\left[R \right]} - 2\,{\hat N}^{-1}\,\hat D^l \hat D_l \hat N +2\,\boldsymbol\kappa\,({\rm\bf K}^l{}_{l})+\tfrac12({\rm\bf K}^l{}_{l})^2 -2\,{\rm\bf k}{}^{l}{\rm\bf k}{}_{l}- \interior{\rm\bf K}{}_{kl}\,\interior{\rm\bf K}{}^{kl} -2\,\mathfrak{e}=0\,.  
\end{eqnarray}

As the constraints equations are always underdetermined (\ref{scal_constr_ell}) may be considered as a second order partial differential equation for both $\Omega$ and $\hat N$, respectively. If it is considered as a PDE for $\Omega$ the characteristic polynomial of (\ref{scal_constr_ell}) reads as 
\begin{equation}
\sigma(x^i,\xi^i)=(\hat n^i \xi_i)^2+\hat \gamma{}^{ij}\,\xi_i\xi_j=h^{ij}\,\xi_i\xi_j\,,
\end{equation}
which is always positive for non-vanishing $\xi_i$, thereby (\ref{scal_constr_ell}) is indeed an elliptic equation for $\Omega$. 

\medskip

If (\ref{scal_constr_ell}) considered to be a PDE for $\hat N$ then, by applying (\ref{trhatK}) and (\ref{trhatK2}), minus one half of the second term can also be written as 
\begin{align}
-{\hat{N}}^{-1}{} & {} \left\{(\mycal{L}_{\hat n} \ln\hat N)\,[\mycal{L}_{\rho} \ln \Omega^2-\hat {D}_k \hat N^k]-\mycal{L}_{\hat n}\,[\mycal{L}_{\rho} \ln \Omega^2-\hat {D}_k \hat N^k]\right\} \nonumber \\ {} & {} { = -{\hat K^l}{}_{l}\,(\mycal{L}_{\hat n} \ln\hat N)+{\hat{N}}^{-1}\mycal{L}_{\hat n}\,[\mycal{L}_{\rho} \ln \Omega^2-\hat {D}_k \hat N^k]}
\end{align}
which, along with the fifth term of (\ref{scal_constr_ell}), {verifies} that the very same equations which is elliptic for $\Omega$ is indeed  parabolic for $\hat N$. {The solubility of this parabolic equation does not allow the vanishing of the mean curvature ${\hat K^l}{}_{l}={\hat{N}}^{-1}{}\,[\mycal{L}_{\rho} \ln \Omega^2-\hat {D}_k \hat N^k]$ on an open domain in $\Sigma_0$ which, in the generic case, is guaranteed by the functional independence of the geometrically distinguished variables.}\,\footnote{ For mathematical conveniences considerations are often restricted to foliations where the mean curvature has a definite sign. For instance, in \cite{bartnik,schwein1,schwein2} (see also \cite{chris}) the positivity of ${\hat K^l}{}_{l}$ was ensured by applying quasiconvex foliations.}

\medskip

By combining all the above observations with Proposition \ref{mom_contsr} we get 
\begin{theorem}
The Hamiltonian and momentum constraints can be given by the system (\ref{scal_constr_ell}), (\ref{par_const2}) and (\ref{ort_const2}). When writing these equations out in coordinates $(\rho,x^3,x^4)$ adopted to the foliation $\mycal{S}_{\rho}$ and the vector field $\rho^i$ on $\Sigma_0$ they can be seen to comprise either an elliptic-hyperbolic or a parabolic-hyperbolic system according to the choice we made for the dependent variable in (\ref{scal_constr_ell}). In either case all the coefficients and source terms smoothly depend on the chosen geometrically distinguished variables.  
\end{theorem} 

Note that although the adopted coordinates $(\tau,\rho,x^3,x^4)$, as they were specified in Section \ref{prelim}, are very rigid in certain extent the spatial coordinates $(\rho,x^3,x^4)$ can be chosen at our will on the initial data surface. In particular, they could be chosen to fit them to the inverse mean curvature flow or to other type of requirements according to the needs of a specific problem to be solved.

\subsection{Solving the constraints}\label{BIVP} 

The solubility of the yielded elliptic-hyperbolic or parabolic-hyperbolic systems, comprised by (\ref{scal_constr_ell}) and (\ref{constr_hyp}), is of fundamental interest. 

In turning to this problem it is rewarding to inspect first the functional form of the coefficients and the source terms of equations  (\ref{scal_constr_ell}) and (\ref{constr_hyp}). All of these can be expressed in terms of our basic variables $\Omega, \gamma_{AB}, \hat N, \hat N^A$ on $\Sigma_0$. 

\medskip

It gets clear after a short inspection that if (\ref{scal_constr_ell}) is considered to be an elliptic equation for the conformal factor $\Omega$ the elliptic-hyperbolic system comprised by (\ref{scal_constr_ell}) and (\ref{constr_hyp}) is ill-posed as these equations need to be solved simultaneously on $\Sigma_0$. { This, however, cannot be done in a consistent way as}  (\ref{scal_constr_ell}) could only be solved as {an elliptic} problem whereas the hyperbolic part should be solved as an initial value problem with initial values specified at some of the two-surfaces $\mycal{S}_\rho$ foliating $\Sigma_0$. {Note also that this elliptic-hyperbolic problem differs significantly from the conventional evolutionary problems as in the latter case the elliptic equations are expected to be solved on the time level surfaces whereas the hyperbolic ones on the entire base manifold. As opposed to this in our case both the elliptic and the 
hyperbolic equations should have to be solved simultaneously on the initial data surface.} 

\medskip

As opposed to the elliptic-hyperbolic system, whenever (\ref{scal_constr_ell}) is considered to be a parabolic equation for the lapse function $\hat N$ the parabolic-hyperbolic system comprised by (\ref{scal_constr_ell}) and (\ref{constr_hyp}) can be solved as an initial value problem with initial data specified at some $\mycal{S}_0\subset\Sigma_0$ for the variables $\hat N, \mycal{L}_n\Omega, \mycal{L}_n \hat N^{A}$, provided that the foliating two-surfaces are orientable compact manifolds with no boundary in $M$.\,\footnote{It is important to be emphasized that whenever the $\mycal{S}_{\tau,\rho}$ surfaces are compact but with boundary or non-compact but infinite (\ref{scal_constr_ell}) and (\ref{constr_hyp}) may also turn to be soluble. Note, however, that then the very same parabolic-hyperbolic system of equations need to be solved as a boundary-initial value problem. In the compact case the suitable boundary values have to be prescribed at the boundaries whereas if the surfaces $\mycal{S}_{\tau,\rho}$ 
are infinite certain fall off conditions have to be specified at the asymptotic regions. Even though the outlined approach may turn to be viable the realization of it is out of the scope of the present paper.}

\medskip

In solving this initial value problem the other dependent variables, i.e.~$\Omega, \gamma_{AB}$, $\mycal{L}_n\,\hat N, \hat N^A, \mycal{L}_n\,\gamma_{AB}$ are  all  freely specifiable on $\Sigma_0$. They, along with the constrained ones determine the full set of initial data for the evolution equations (\ref{ham_1+(1+2)})-(\ref{evol_1+(1+2)}). 

\medskip

In getting some insight concerning the desired well-posedness of the initial value problem associated with our parabolic-hyperbolic system, (\ref{scal_constr_ell}) and (\ref{constr_hyp}), recall first that the initial value problem for the hyperbolic system is always well-posed \cite{helmut, helmut&alan}. In addition, the separate parabolic systems---at least in certain special cases {with quasiconvex foliations}---are known to be well-posed possessing globally regular solutions \cite{bartnik,schwein1,schwein2}. Note also that the principal parts of the parabolic and hyperbolic equations are independent whereby the well-posedness of the coupled system {is} also {expected to} follow.

\medskip

In summarizing the above observations we have

\begin{corollary}
The Hamiltonian and momentum constraints can be given by the parabolic-hyperbolic system comprised by (\ref{scal_constr_ell}) and (\ref{constr_hyp}). Provided that the two-surfaces foliating the spacetime are orientable compact manifolds with no boundary in $M$ this parabolic-hyperbolic system can be solved as an initial value problem on the three-dimensional initial data surface $\Sigma_0$ for 
\begin{equation}
\hat N, \mycal{L}_n\Omega, \mycal{L}_n \hat N^{A}
\end{equation}
once a sufficiently regular choice for the variables
\begin{equation}\label{list-free_a}
\Omega, \gamma_{AB}, \mycal{L}_n\,\hat N, \hat N^A, \mycal{L}_n\,\gamma_{AB}
\end{equation}
had been made throughout $\Sigma_0$. We also have the freedom to choose initial data to the parabolic-hyperbolic system on one of the two-surfaces foliating $\Sigma_0$.
\end{corollary}

\section{The evolutionary system}\label{evolutionary} 
\setcounter{equation}{0}

This section is to spell out the evolutionary system consisting of (\ref{ham_1+(1+2)})-(\ref{evol_1+(1+2)}) by making use of our geometrically distinguished variables. 

\subsection{The equations} \label{eveq}

Recall first that the $1+2$ `Hamiltonian and momentum' constraints, (\ref{ham_1+(1+2)}) and (\ref{mom_1+(1+2)}), read as 
\begin{eqnarray}  
\hat E^{{}^{(\mathcal{H})}} {}&\hskip-.4cm=&\hskip-.2cm{} \tfrac12\,\{ -\hat R + ({{\hat K}{}^{l}}{}_{l})^2 - \hat K_{kl}\hat K^{kl} - 2\,\hat{\mathfrak{e}} \}=0\,,\label{ham_1+(1+2)_detailed} \\
\hat E^{{}^{(\mathcal{M})}}_i {}&\hskip-.4cm=&\hskip-.2cm{} \hat D^l {{\hat K}}{}_{li} - \hat D_i {{\hat K}{}^{l}}{}_{l} - \hat{\mathfrak{p}}_{i}=0\,.\label{mom_1+(1+2)_detailed}
\end{eqnarray}

In determining the rest of the evolutionary system it is advantageous to take the trace and trace free parts of (\ref{evol_1+(1+2)}) which, along with the use of some algebra, yields the relations
\begin{eqnarray}\label{evol_1+(1+2)_tr}
\hat \gamma^{kl}\,\hat E^{{}^{(\mathcal{EVOL})}}_{kl} {}&\hskip-.2cm = &\hskip-.2cm{} \hat R + \mycal{L}_{\hat n}({\hat K}{}^l{}_{l})-({\hat K}{}^l{}_{l})^2+2\,\hat K_{kl}\hat K^{kl} + {\hat{N}}^{-1}\,(\hat D^l \hat D_l \hat N) - [\hat{\mathfrak{S}}_{kl}\,\hat \gamma^{kl}-2\,\hat{\mathfrak{e}}] \nonumber \\
{}&\hskip-.2cm=&\hskip-.2cm{} -2\,\hat E^{{}^{(\mathcal{H})}} - \hat{\mathfrak{S}}_{kl}\,\hat \gamma^{kl}+\mycal{L}_{\hat n}({\hat K}{}^l{}_{l}) + \hat K_{kl}\hat K^{kl}+ {\hat{N}}^{-1}\,(\hat D^l \hat D_l \hat N)
\end{eqnarray} 
\begin{eqnarray}\label{evol_1+(1+2)_trf}
&& \hskip-0.7cm \hat R_{ij}-\tfrac12\,\hat R\,\hat \gamma_{ij}= [\hat E^{{}^{(\mathcal{EVOL})}}_{ij}-\tfrac12\,\hat \gamma_{ij}\,(\hat \gamma^{kl}\,\hat E^{{}^{(\mathcal{EVOL})}}_{kl})] + \mycal{L}_{\hat n}{\hat K}_{ij}+({\hat K}{}^l{}_{l})\,{\hat K}_{ij}-2\,\hat K_{il}{\hat K}{}^l{}_{j} \nonumber \\
&& \hskip-0.7cm\phantom{\hat R_{ij}}+ {\hat{N}}^{-1}\,(\hat D_i \hat D_j \hat N)-\tfrac12\,\hat \gamma_{ij}\left\{\mycal{L}_{\hat n}({\hat K}{}^l{}_{l})+({\hat K}{}^l{}_{l})^2 +{\hat{N}}^{-1}\,(\hat D^l \hat D_l \hat N)\right\} + [\hat{\mathfrak{S}}_{ij}-\tfrac12\,\hat \gamma_{ij}\,(\hat{\mathfrak{S}}_{kl}\,\hat \gamma^{kl})] \nonumber\\
&& \hskip-0.7cm\phantom{\hat R_{ij}}=\Pi^{kl}{}_{ij}\,\left[\hat E^{{}^{(\mathcal{EVOL})}}_{kl} + \mycal{L}_{\hat n}{\hat K}_{kl}+ ({\hat K}{}^m{}_{m})\,{\hat K}_{kl}+ {\hat{N}}^{-1}\,(\hat D_k \hat D_l \hat N) + \hat{\mathfrak{S}}_{kl}\right]\,.
\end{eqnarray}
Thereby, whenever (\ref{ham_1+(1+2)_detailed}) holds the $1+2$ `evolution equation' $\hat E^{{}^{(\mathcal{EVOL})}}_{ij}=0$ can be seen to be equivalent to the pair of equations 
\begin{eqnarray}\label{evol_1+(1+2)_tr_1}
- \hat{\mathfrak{S}}_{kl}\,\hat \gamma^{kl}+\mycal{L}_{\hat n}({\hat K}{}^l{}_{l}) + \hat K_{kl}\hat K^{kl}+ \hat D^l \hat D_l (\ln \hat N)+(\hat D^l \ln \hat N)(\hat D_l \ln \hat N)=0
\end{eqnarray} 
\begin{eqnarray}\label{evol_1+(1+2)_trf_1}
\Pi^{kl}{}_{ij}\,\left[\mycal{L}_{\hat n}{\hat K}_{kl}+ ({\hat K}{}^m{}_{m})\,{\hat K}_{kl}+ \hat D_k \hat D_l (\ln \hat N)+(\hat D_k \ln \hat N)(\hat D_l \ln \hat N) + \hat{\mathfrak{S}}_{kl}\right]=0\,.
\end{eqnarray}

By inspecting the terms involved in the evolutionary system formed by (\ref{ham_1+(1+2)_detailed}), (\ref{mom_1+(1+2)_detailed}), (\ref{evol_1+(1+2)_tr_1}) and (\ref{evol_1+(1+2)_trf_1}) it gets to be immediately clear that in order to uncover all the hidden second order `time' derivatives the source terms $\hat{\mathfrak{e}}=\hat n^k \hat n^l\,{}^{{}^{(3)}}\hskip-1mm \mycal{G}_{kl}$, $\hat{\mathfrak{p}}_{i}={{\hat \gamma}^{k}}{}_{i}\,\hat n^l\, {}^{{}^{(3)}}\hskip-1mm \mycal{G}_{kl}$, $\hat \gamma^{kl}\,\hat{\mathfrak{S}}_{kl}=\hat \gamma^{kl}\,[{{\hat \gamma}^{p}}{}_{k} {{\hat \gamma}^{q}}{}_{l}\,\,{}^{{}^{(3)}}\hskip-1mm \mycal{G}_{pq}]=\hat \gamma^{kl}\,{}^{{}^{(3)}}\hskip-1mm \mycal{G}_{kl}$ and $\Pi^{kl}{}_{ij}\hat{\mathfrak{S}}_{kl}=\Pi^{kl}{}_{ij}\,[{{\hat \gamma}^{p}}{}_{k} {{\hat \gamma}^{q}}{}_{l}\,{}^{{}^{(3)}}\hskip-1mm \mycal{G}_{pq}]=\Pi^{kl}{}_{ij}\,{}^{{}^{(3)}}\hskip-1mm \mycal{G}_{kl}$ have to be evaluated. It worth keeping in mind that, in virtue of (\ref{int_evol_1+3_ch}), 
these source terms---as 
opposed to ${\mathfrak{e}}$, ${\mathfrak{p}}_
{i}$ and ${\mathfrak{S}}_{ij}$ given by (\ref{1+3source})---do not vanish even if considerations are restricted to the pure vacuum Einstein's equations. 

\medskip

A detailed derivation of these source terms can be found in the appendix. In  order to keep the expressions to be reasonably compact hereafter in most of the cases only the principal parts of the expressions will be spelled out explicitly in terms of the chosen {geometrically distinguished} variables $\Omega, \gamma_{ij}, \hat N, \hat N^i;\nu$.

\subsection{The principal parts of the evolution equations} \label{princeveq}

To start off consider first the $1+2$ ``Hamiltonian constraint'' (\ref{ham_1+(1+2)_detailed}) in which, in addition to $\hat{\mathfrak{e}}$ only the scalar curvature $\hat{R}$ contains implicit second order derivatives of the conformal factor $\Omega$ and the metric $\gamma_{ij}$. By taking (\ref{ehat}) and (\ref{hatextcurv2}) into account the principal part of (\ref{ham_1+(1+2)_detailed}) can be seen to take the form\,\footnote{It is worth mentioning that the term ${}^{{}^{(\gamma)}}\hskip-1mm R$ contains second order derivatives of $\gamma_{ij}$---which are tangential to the level surfaces $\mycal{S}_{\tau,\rho}$---as
\begin{equation}
{}^{{}^{(\gamma)}}\hskip-1mm R= \gamma^{kl}\gamma^{pq}(\partial_k\partial_p\gamma_{lq}-\partial_k\partial_l\gamma_{pq}) + \{\rm lower\ order\ terms \}\,.
\end{equation}
} 
\begin{align}\label{Evol_nn} 
-2\,\mycal{L}^2_n \,\Omega + \hat{D}^l\hat{D}_l\, \Omega + \Omega\,\hat D{}^l\hat D_l (\ln  N) - \tfrac12{}^{{}^{(\gamma)}}\hskip-1mm R\,{\Omega}^{-1} + \{\rm lower\ order\ terms \} =0  \,.
\end{align}

Similarly, by taking into account the relations (\ref{hatK}) and (\ref{trhatK2}) we get that 
\begin{equation}
\hat D^l \hat K_{li}=\tfrac12\Omega^2\left[\hat D^l (\mycal{L}_{\hat n}\gamma_{li}) + (\mycal{L}_{\hat n}\gamma_{li})\hat D^l(\ln \Omega^2) \right]
+\tfrac12 \hat D_i[\mycal{L}_{\hat n}(\ln \Omega^2)]
\end{equation}
\begin{equation}
\hat D_i \hat K^l{}_{l}={\hat{N}}^{-1}\left[\hat D_i [\mycal{L}_{\rho}(\ln \Omega^2)]- \hat D_i(\hat D_l \hat N^l)  - \hat D_i(\ln \hat N)\left( \mycal{L}_{\rho}(\ln \Omega^2)-(\hat D_l \hat N^l)\right)  \right]\,.
\end{equation}
These latter relations, along with (\ref{phat}), impl{y} that the principal part of the $1+2$ ``momentum constraint'' (\ref{mom_1+(1+2)_detailed}) takes the form 
\begin{align}\label{Evol_pn}
{(2\hat N)}^{-1}{\Omega^2}\,\gamma_{il}(\mycal{L}^2_{n}\hat N^l){}&+{\hat{N}}^{-1}\,\hat D_i(\hat D_l \hat N^l)+ \tfrac12\Omega^2\,\hat D^l [\mycal{L}_{\hat n}\gamma_{li}]- \hat D_i [\mycal{L}_{\hat n}(\ln  N)]\\ +{}&\,{\Omega}^{-1}\,\hat D_i[\mycal{L}_{\hat n}\,\Omega]- 2\,{(\hat N\Omega)}^{-1}\hat D_i [\mycal{L}_{\rho}\,\Omega]  + \{\rm lower\ order\ terms \} =0\,. \nonumber
\end{align}

Analogously, using the expression we have for the term $\hat \gamma^{kl}\,\hat{\mathfrak{S}}_{kl}$, given by (\ref{1_ghat_sigmahat}), and applying the relation 
\begin{equation}
\mycal{L}_{\hat n} \hat K^l{}_{l}= {\hat{N}}^{-1}\left\{ \mycal{L}_{\hat n}\mycal{L}_{\rho}(\ln \Omega^2) - \mycal{L}_{\hat n}(\hat D_l \hat N^l) + \mycal{L}_{\hat n}(\ln \hat N) \left[\mycal{L}_{\rho}(\ln \Omega^2)-(\hat D_l \hat N^l)\right]\right\}\,
\end{equation}
the principal part of the trace-equation (\ref{evol_1+(1+2)_tr_1}) can be given as 
\begin{align}\label{Evol_trace0}
-2\,\mycal{L}^2_{n}(\ln \hat N){}&-2\,\Omega^{-1}\mycal{L}^2_{n}\,\Omega+ 2\,\mycal{L}^2_{\hat n}(\ln \hat N) + \hat D^l \hat D_l(\ln N +\ln \hat N) \\ {}&+2\,{(\hat N\Omega)}^{-1}\mycal{L}_{\hat n}\mycal{L}_{\rho}\,\Omega-{\hat{N}}^{-1} \mycal{L}_{\hat n}(\hat D_l \hat N^l)+ \{\rm lower\ order\ terms \} =0\,. \nonumber
\end{align}
In order to get a pure evolution equation for $\hat N$ (\ref{Evol_trace0}) may be combined with (\ref{Evol_nn}) which yields 
\begin{align}\label{Evol_trace}
-2\,\mycal{L}^2_{n}{}&(\ln \hat N)+ 2\,\mycal{L}^2_{\hat n}(\ln \hat N) + \hat D^l \hat D_l(\ln \hat N)+2\,{(\hat N\Omega)}^{-1}\mycal{L}_{\hat n}\mycal{L}_{\rho}\,\Omega \\ {}&-{\hat{N}}^{-1} \mycal{L}_{\hat n}(\hat D_l \hat N^l)- {\Omega}^{-1}\,\hat{D}^l\hat{D}_l\, \Omega+ \tfrac12{}^{{}^{(\gamma)}}\hskip-1mm R\,{\Omega}^{-2} + \{\rm lower\ order\ terms \} =0\,. \nonumber 
\end{align}

Finally, in determining the principal part of (\ref{evol_1+(1+2)_trf_1}) notice first that 
\begin{equation}
\mycal{L}_{\hat n} \hat K_{ij}= \tfrac12\Omega^2\left\{ \left[\mycal{L}^2_{\hat n}(\ln \Omega^2) + \left(\mycal{L}_{\hat n}(\ln \Omega^2)\right)^2\right]\,\gamma_{ij} + 
\mycal{L}^2_{\hat n}\,\gamma_{ij} + 2 \,(\mycal{L}_{\hat n}\gamma_{ij})\mycal{L}_{\hat n}(\ln \Omega^2)\right\}\,,
\end{equation}
from which, in virtue of (\ref{PI_negyzet}) and (\ref{PI2}) we get that 
\begin{align}
\Pi^{kl}{}_{ij}\,\mycal{L}_{\hat n} \hat K_{kl}{}&= \tfrac12\,\Omega^2\,\Pi^{kl}{}_{ij}\left[ \mycal{L}^2_{\hat n}\gamma_{kl} + 2 \,(\mycal{L}_{\hat n}\gamma_{ij})\mycal{L}_{\hat n}(\ln \Omega^2)\right] \\ {}&=\tfrac12\,\Omega^2\left[ \gamma^k{}_i\gamma^l{}_j\mycal{L}^2_{\hat n}\gamma_{kl}-\tfrac12\,\gamma_{ij} (\gamma^{kl}\mycal{L}^2_{\hat n}\gamma_{kl})+ 2\,\Pi^{kl}{}_{ij}\,(\mycal{L}_{\hat n}\gamma_{kl})\mycal{L}_{\hat n}(\ln \Omega^2)\right]\,. \nonumber
\end{align}
Taking then into account (\ref{xi}), with $\eta^a=\rho^a$, we get $\gamma^{kl}(\mycal{L}_{\hat n}\gamma_{kl})= -{\hat{N}}^{-1}(\mathbb{D}_l \hat N^l)$ which, along with
\begin{align}\label{g_Lie2_g}
\gamma^{kl}(\mycal{L}^2_{\hat n}\gamma_{kl}){}&=\mycal{L}_{\hat n}\left(\gamma^{kl}(\mycal{L}_{\hat n}\gamma_{kl})\right)-(\mycal{L}_{\hat n}\gamma^{kl})(\mycal{L}_{\hat n}\gamma_{kl})  \\ {}&= -2{\hat{N}}^{-1}\mycal{L}_{\hat n}(\mathbb{D}_l \hat N^l)+2{\hat{N}}^{-1}\mycal{L}_{\hat n}(\ln \hat N)(\mathbb{D}_l \hat N^l)+ \gamma^{kp} \gamma^{lq} (\mycal{L}_{\hat n}\gamma_{kl}) (\mycal{L}_{\hat n}\gamma_{pq})\,,\nonumber
\end{align}
verifies that 
\begin{align}\label{PILieDnhatKhat}
\hskip-0.3cm\Pi^{kl}{}_{ij}\,\mycal{L}_{\hat n} \hat K_{kl}=\tfrac12\,\Omega^2\,\left[ \gamma^k{}_i\gamma^l{}_j\mycal{L}^2_{\hat n}\gamma_{kl}+{\hat N}^{-1}\gamma_{ij}\mycal{L}_{\hat n}(\mathbb{D}_l \hat N^l)\right] + \{\rm lower\ order\ terms \} \,. 
\end{align} 
Thus, by combining (\ref{evol_1+(1+2)_trf_1}), (\ref{PILieDnhatKhat}) and (\ref{PI_sigmahat}) the principal part of (\ref{evol_1+(1+2)_trf_1}) can be given as 
\begin{align}\label{princ_PI}
-\tfrac12\,\Omega^2\, \left[\gamma^k{}_i\gamma^l{}_j(\mycal{L}^2_{n}\gamma_{kl}  -\mycal{L}^2_{\hat n}\gamma_{kl}) -{\hat N}^{-1}\gamma_{ij}\mycal{L}_{\hat n}(\mathbb{D}_l \hat N^l)\right]  {}&+ \Pi^{kl}{}_{ij}\left[ \hat D_k \hat D_l(\ln N +\ln \hat N) \right] \nonumber \\ {}& + \{\rm lower\ order\ terms \} \,. 
\end{align}

\section{Solving the mixed hyperbolic-hyperbolic system} \label{evolutionary_hyp_hyp}

Having the full reduced set of Einstein's equations some remarks are in order now. Note first that by demanding the three-dimensional shift vector to vanish and by determining all the coordinates as described in Section \ref{prelim} the applied kinematic setup became very rigid. In particular, it does not allow the application, e.g.~of the generalized harmonic coordinates (see Section 2 in \cite{helmut}) to put our evolution equations into a strongly hyperbolic system to which well-posedness results are known to apply. 

\medskip

Another consequence of demanding the three-dimensional shift vector to vanish is that we left one (and only one) of the gauge source functions to be unspecified. This is in accordance with the fact that no evolution equation applies to the auxiliary variable $\nu$ relating the tree- and two-dimensional lapses $N$ and $\hat N$ as $N=\nu\,\hat N$. It is important to keep in mind that this freedom may become useful in numerical simulations. For instance, by suitably evolving $\nu$---as the `$\beta$' function was evolved in \cite{CSPIR}---one could adjust the time-slices to acquire the largest possible Cauchy development for a given initial data specification. 

\medskip

Note also that by following the conventional $1+3$ decomposition based routines to get a solution to (\ref{geom}) first the coupled parabolic-hyperbolic system (\ref{scal_constr_ell}) and (\ref{constr_hyp}) should be solved for the constraints on the initial data surface $\Sigma_0$. { Then the evolutionary system (\ref{Evol_nn}), (\ref{Evol_pn}), (\ref{Evol_trace}),  (\ref{princ_PI})} is expected to be {solved.} This {process} presumes that the latter equations can be shown to possess a well-posed initial value formulation which may be the case although no attempt will be made to investigate this problem here. Instead a more tempting approach is proposed below. 

\medskip

Indeed, the rest of this section is to justify that by combining the {first order symmetric} hyperbolic part of the constraint equations with the manifestly strongly hyperbolic members of the $1+2$ reduced equations a mixed hyperbolic-hyperbolic system may be formed to which a well-posed initial value problem {expected to apply}.  

\medskip 

For the sake of simplicity let us assume that the two-surfaces foliating the spacetime are orientable compact manifolds with no boundary in $M$. Recall also that once the variables $\hat N, \hat N^A, \Omega, \gamma_{AB}, \mycal{L}_n\,\hat N, \mycal{L}_n\,\gamma_{AB}$ are known on any of the $\Sigma_\tau$ time-level surfaces the {hyperbolic} system comprised by (\ref{par_const2}) and (\ref{ort_const2}) can be solved as an initial value problem for $ \mycal{L}_n\Omega, \mycal{L}_n \hat N^{A}$ provided that initial data are given for $\mycal{L}_n\Omega, \mycal{L}_n \hat N^{A}$ on one of the two-surfaces (denote it by $\mycal{S}$) foliating $\Sigma_\tau$. Knowing these Lie derivatives throughout $\Sigma_\tau$ the values of $\Omega$ and $\hat N^{A}$, on the succeeding time-level surface can be determined by (a forward in time) integration  along the $\tau$-coordinate lines. Notice also that the initial data for $\mycal{L}_n\Omega, \mycal{L}_n \hat N^{A}$ on the intersection of the world-tube (denote it 
by $\mycal{W}_{\mycal{S}
}$)---yielded by the Lie transport of $\mycal{S}$ along the time evolution vector field {$\tau^a$}---and the succeeding time-level surface can be determined by making use of the evolution equations (\ref{Evol_nn}) and (\ref{Evol_pn}) on $\mycal{W}_{\mycal{S}}$. 

\medskip 

In order to be able to update all the basic variables on the succeeding time slice, in addition, to all {the} above equations (\ref{Evol_trace}) and (\ref{princ_PI}) also have to be solved for $\hat N$ and $\gamma_{AB}$. Note that as (\ref{Evol_trace}) and (\ref{princ_PI}) comprise a manifestly strongly hyperbolic system {(see \cite{racz_tdfd} for more details)} such that neither of the involved source terms refers to objects {other} than the tangential (to the $\Sigma_\tau$ time-level surfaces) derivatives of the basic variables, $\hat N, \hat N^A, \Omega, \gamma_{AB}; \nu$. By solving these evolution equations all the variables $\hat N, \hat N^A, \Omega, \gamma_{AB}, \mycal{L}_n\,\hat N, \mycal{L}_n\,\gamma_{AB}$ get to be updated on the succeeding time-level surface which ensures then that by the repetition of the process described in the last two paragraphs we get a solution for $\hat N, \hat N^A, \Omega, \gamma_{AB}$ throughout $M$. 

\medskip

By combining the above observations we get 
\begin{theorem}
Assume that {$M=\mathbb{R}^2\times {\mycal{S}}^2$ and that} a function $\nu: M\rightarrow \mathbb{R}$ had been chosen and that suitable initial data satisfying the $1+3$ Hamiltonian and momentum constraints had been specified on the initial data surface $\Sigma_0$. Denote by $\mycal{S}_0$ one of the members of the two-surfaces foliating $\Sigma_0$ and by $\mycal{W}_{\mycal{S}_0}$ the world-tube yielded by the Lie transport of $\mycal{S}_0$ along the time evolution vector field $\tau^a$. Then by solving the first order symmetric hyperbolic system, comprised by (\ref{par_const2}) and (\ref{ort_const2}), for $ \mycal{L}_n\Omega, \mycal{L}_n \hat N^{A}$ on each of the time-level surfaces---such that the initial data for these equations are determined by making use of the evolution equations (\ref{Evol_nn}) and (\ref{Evol_pn}) along the world-tube $\mycal{W}_{\mycal{S}_0}$---and also by solving the strongly hyperbolic evolution equations (\ref{Evol_trace}) and (\ref{princ_PI}) a solution {to the mixed hyperbolic-
hyperbolic system comprised by (\ref{par_const2}), (\ref{ort_const2}), (\ref{Evol_trace}) and (\ref{princ_PI})} for the geometrically distinguished variables $\hat N,\hat N^A,\Omega,\gamma_{AB}$ can be determined {throughout $M$}. 
\end{theorem} 

\medskip

Assume that we have a solution to the above mixed hyperbolic-hyperbolic system  which, along with $\nu$, determines a spacetime metric of the form (\ref{pref}). It is of obvious interest to know whether this solution is also a solution to the full set of Einstein's equations. To see that the answer is in the affirmative we could proceed as follows. 

\medskip 

In principle the Bianchi identity provides us the needed relations of the constraint and evolution equations. In deriving them start by applying (3.17) and (3.18) of \cite{racz_geom_det} with a simultaneous substitution of $\mathfrak{e}$, $\mathfrak{p}_a$ and $\mathfrak{S}_{ab}$ by $E^{{}^{(\mathcal{H})}}$, $-E^{{}^{(\mathcal{M})}}_a$ and $E^{{}^{(\mathcal{EVOL})}}_{ab}\hskip-.1cm{} + h_{ab} \,E^{{}^{(\mathcal{H})}}$, respectively, which yields the relations 
\begin{align} 
\mycal{L}_n\,E^{{}^{(\mathcal{H})}} - D^e E^{{}^{(\mathcal{M})}}_e +\left[ 2\,E^{{}^{(\mathcal{H})}}\,({K^e}_e) -2\,(\dot n^e\,E^{{}^{(\mathcal{M})}}_e) +K^{ab}\,E^{{}^{(\mathcal{EVOL})}}_{ab}\hskip-.1cm{} \,\right] {}&= 0\, \label{cons_law1} \\ 
\mycal{L}_n\,E^{{}^{(\mathcal{M})}}_b - D^aE^{{}^{(\mathcal{EVOL})}}_{ab}\hskip-.1cm{} - D_{b}E^{{}^{(\mathcal{H})}} - \left[E^{{}^{(\mathcal{EVOL})}}_{ab}\hskip-.1cm{}\,\dot n^a - ({K^e}_e)\,E^{{}^{(\mathcal{M})}}_b+2\, E^{{}^{(\mathcal{H})}}\,\dot n_b   \,\right] {}& = 0\,. \label{cons_law2}
\end{align}

\medskip

In addition, we may also decompose $E^{{}^{(\mathcal{EVOL})}}_{ab}$ as 
\begin{align}
E^{{}^{(\mathcal{EVOL})}}_{ab}\hskip-.1cm{}=\hat E^{{}^{(\mathcal{H})}}\hat n_a\hat n_b + [ \hat n_a\hat E^{{}^{(\mathcal{M})}}_b\hskip-.1cm{} +\hat n_b\hat E^{{}^{(\mathcal{M})}}_a ] + (\hat E^{{}^{(\mathcal{EVOL})}}_{ab}\hskip-.1cm{} + \hat\gamma_{ab}\hat E^{{}^{(\mathcal{H})}} )\,.
\end{align}
Note also that (\ref{evol_1+(1+2)_tr_1}) and (\ref{evol_1+(1+2)_trf_1})---in virtue of (\ref{evol_1+(1+2)_tr}) they can be seen to be equivalent to $\Pi^{ij}{}_{ab}\hat E^{{}^{(\mathcal{EVOL})}}_{ij}\hskip-.1cm{}=0$ and $\hat\gamma^{ij}\hat E^{{}^{(\mathcal{EVOL})}}_{ij}\hskip-.1cm{}+2\,\hat E^{{}^{(\mathcal{H})}}=0$---can be used to verify 
\begin{align}
\hat E^{{}^{(\mathcal{EVOL})}}_{ab}\hskip-.1cm{} + \hat\gamma_{ab}\hat E^{{}^{(\mathcal{H})}}{}&=[\,\Pi^{ij}{}_{ab}+\tfrac12\hat\gamma_{ab}\,\hat\gamma^{ij}\,] \, \hat E^{{}^{(\mathcal{EVOL})}}_{ij}\hskip-.1cm{} + \hat\gamma_{ab}\,\hat E^{{}^{(\mathcal{H})}} \nonumber\\ {}&= \Pi^{ij}{}_{ab}\,\hat E^{{}^{(\mathcal{EVOL})}}_{ij}\hskip-.1cm{} +  \tfrac12\hat\gamma_{ab}\, [\,\hat\gamma^{ij}\hat E^{{}^{(\mathcal{EVOL})}}_{ij}\hskip-.1cm{} + 2\,\hat E^{{}^{(\mathcal{H})}}]=0\,.
\end{align} 
Thus, whenever (\ref{evol_1+(1+2)_tr_1}) and (\ref{evol_1+(1+2)_trf_1}) holds we have 
\begin{align}
E^{{}^{(\mathcal{EVOL})}}_{ab}\hskip-.1cm{}=\hat E^{{}^{(\mathcal{H})}}\hat n_a\hat n_b + [ \hat n_a\hat E^{{}^{(\mathcal{M})}}_b\hskip-.1cm{} +\hat n_b\hat E^{{}^{(\mathcal{M})}}_a ]\,.
\end{align}  
This, along with (\ref{dec_1}), verifies then the relations
\begin{align}
K^{ab}E^{{}^{(\mathcal{EVOL})}}_{ab}\hskip-.1cm{} {}&=\boldsymbol\kappa\,\hat E^{{}^{(\mathcal{H})}}  + 2\,{\rm\bf k}{}^{e}\hat E^{{}^{(\mathcal{M})}}_e \\  
\dot n^a E^{{}^{(\mathcal{EVOL})}}_{ab}\hskip-.1cm{} {}&=[(\hat n_a \dot n^a)\hat E^{{}^{(\mathcal{H})}} + (\dot n^a\,\hat E^{{}^{(\mathcal{M})}}_a )]\hat n_b + (\hat n_a \dot n^a) \hat E^{{}^{(\mathcal{M})}}_b\,.
\end{align} 
By applying (A.9) and (A.10) of \cite{racz_geom_det}, with $\epsilon=1$, we also get 
\begin{align}
{} \hat\gamma^e{}_b D^{a}E^{{}^{(\mathcal{EVOL})}}_{ae}\hskip-.1cm{} {}&=(\hat K^e{}_e) \,\hat E^{{}^{(\mathcal{M})}}_b + \dot {\hat n}_b\, \hat E^{{}^{(\mathcal{H})}} + \mycal{L}_{\hat n}\,\hat E^{{}^{(\mathcal{M})}}_b \\  
{} \hat n^e{}D^{a}E^{{}^{(\mathcal{EVOL})}}_{ae}\hskip-.1cm{}  {} &=(\hat K^e{}_e) \,\hat E^{{}^{(\mathcal{H})}} + \hat D^e \hat E^{{}^{(\mathcal{M})}}_e  + \mycal{L}_{\hat n}\,\hat E^{{}^{(\mathcal{H})}} - 2\,\dot {\hat n}^e  \hat E^{{}^{(\mathcal{M})}}_e \\ 
D_b  E^{{}^{(\mathcal{H})}} {}&= [\hat\gamma^e{}_b + \hat n^e\hat n_b]\, D_e E^{{}^{(\mathcal{H})}} = \hat D_b E^{{}^{(\mathcal{H})}} + {\hat n}_b\, \mycal{L}_{\hat n}\,E^{{}^{(\mathcal{H})}} \,.
\end{align} 

\medskip

Finally, whenever all the relations $\Pi^{ij}{}_{ab}\hat E^{{}^{(\mathcal{EVOL})}}_{ij}\hskip-.1cm{}=0$, $\hat\gamma^{ij}\hat E^{{}^{(\mathcal{EVOL})}}_{ij}\hskip-.1cm{}+2\,\hat E^{{}^{(\mathcal{H})}}=0$ and $E^{{}^{(\mathcal{M})}}_b\hskip-.1cm{}=0$ hold we get from (\ref{cons_law1}) 
\begin{align} 
\mycal{L}_n\,E^{{}^{(\mathcal{H})}} + 2\,({K^e}_e)\,E^{{}^{(\mathcal{H})}}+ \boldsymbol\kappa\,\hat E^{{}^{(\mathcal{H})}}  + 2\,{\rm\bf k}{}^{e}\hat E^{{}^{(\mathcal{M})}}_e {}&= 0\, \label{cons_law1_2} 
\end{align}
whereas the parallel and orthogonal parts of (\ref{cons_law2}), respectively, yield the following two relations
\begin{align} 
\hskip-0.3cm[(\hat K^e{}_e) \,\hat E^{{}^{(\mathcal{M})}}_b + \dot {\hat n}_b\, \hat E^{{}^{(\mathcal{H})}} + \mycal{L}_{\hat n}\,\hat E^{{}^{(\mathcal{M})}}_b ]+ \hat D_b E^{{}^{(\mathcal{H})}}  + (\hat n_a \dot n^a) \hat E^{{}^{(\mathcal{M})}}_b {}&+ 2\,{} \hat\gamma_{eb}\, \dot n^e\, E^{{}^{(\mathcal{H})}}=0  \label{cons_law2_2a} \\
[(\hat K^e{}_e) \,\hat E^{{}^{(\mathcal{H})}} + \hat D^e \hat E^{{}^{(\mathcal{M})}}_e  + \mycal{L}_{\hat n}\,\hat E^{{}^{(\mathcal{H})}} - 2\,\dot {\hat n}^e  \hat E^{{}^{(\mathcal{M})}}_e] + \mycal{L}_{\hat n}\,E^{{}^{(\mathcal{H})}} {}&+ [(\hat n_a \dot n^a)\hat E^{{}^{(\mathcal{H})}} + (\dot n^a\,\hat E^{{}^{(\mathcal{M})}}_a)] \nonumber \\ {}&+ 2\,(\hat n_a \dot n^a) E^{{}^{(\mathcal{H})}} =0 \,. \label{cons_law2_2b}
\end{align}

\medskip

It follows then that (\ref{cons_law1_2}), (\ref{cons_law2_2a}) and (\ref{cons_law2_2b}) {comprise a mixed hyperbolic-hyperbolic system that} possess{es} the identically zero solution for $E^{{}^{(\mathcal{H})}}, \hat E^{{}^{(\mathcal{H})}}$ and $\hat E^{{}^{(\mathcal{M})}}_a$ provided that the vanishing of $E^{{}^{(\mathcal{H})}}$ on $\Sigma_0$ and the vanishing of $\hat E^{{}^{(\mathcal{H})}}$ and $\hat E^{{}^{(\mathcal{M})}}_a$ on the world-tube $\mycal{W}_{\mycal{S}_0}$ are guaranteed. 

\medskip

What we have can be summarized as 
\begin{theorem}\label{hypTh2}
Solutions to the mixed hyperbolic-hyperbolic system comprised by (\ref{par_const2}), (\ref{ort_const2}), (\ref{Evol_trace}) and (\ref{princ_PI}) are also solutions to the full set of Einstein's equations (\ref{geom}) provided that (\ref{ort_const}) holds on the initial data surface $\Sigma_0$, and (\ref{Evol_nn}) and (\ref{Evol_pn}) are satisfied along the world-tube $\mycal{W}_{\mycal{S}_0}$. 
\end{theorem}

Note that whenever the foliating two-surfaces are compact without boundary in the spacetime and a regular origin exist---this is the location  {on the succeeding time-slices} where the foliating two-surfaces smoothly reduce to a point---in addition to solving the mixed hyperbolic-hyperbolic system it suffices merely to solve the $1+3$ Hamiltonian constraint on the initial data surface. To see this note that (\ref{Evol_nn}) and (\ref{Evol_pn}) always hold at the world-line of the origin provided that the spacetime geometry is regular through the origin, i.e.~the existence of defects such as a conical singularity are excluded. Indeed, the spacetime geometry is regular at an origin signified by $\rho=\rho_*$ if $\hat N, \Omega$ and $\hat N^A$ can be given in its neighborhood as 
\begin{align}\label{origin} 
\hat N=1+  (\rho-\rho_*)^2\,\hat N{}_{(2)}, \ \Omega=(\rho-\rho_*)+ (\rho-\rho_*)^3\,\Omega{}_{(3)} \ \ {\rm and} \ \ \hat N^A= (\rho-\rho_*)\,\hat N^A_{(1)}\,,
\end{align}
where the terms $\hat N{}_{(2)}, \Omega{}_{(3)}$ and $\hat N^A_{(1)}$ smoothly depends on the coordinates $(\tau,\rho,x^3,x^4)$. By substituting the relations listed in (\ref{origin}) into (\ref{Evol_nn}) and (\ref{Evol_pn}) and taking the limit $\rho\rightarrow\rho_*$ (\ref{Evol_nn}) and (\ref{Evol_pn}) can be seen to hold {automatically} along the world-line {$\mycal{W}_{\rho_*}$} of the origin. In virtue of the last two relations of (\ref{origin}), both of the terms $\mycal{L}_n\Omega$ and $\mycal{L}_n \hat N^{A}$ vanish there, i.e.~the initial data for the first order symmetric hyperbolic system comprised by (\ref{par_const2}) and  (\ref{ort_const2}) at $\rho=\rho_*$ has to be trivial. 

\medskip

In summarizing we have  
\begin{corollary}\label{hypTh3}
Assume that a world line {$\mycal{W}_{\rho_*}$} representing an origin exist such that the spacetime geometry is regular at {$\mycal{W}_{\rho_*}$}. Then $\mycal{L}_n\Omega=0$ and $\mycal{L}_n \hat N^{A}=0$ hold along {$\mycal{W}_{\rho_*}$}, and the corresponding solutions to the mixed hyperbolic-hyperbolic system comprised by (\ref{par_const2}), (\ref{ort_const2}), (\ref{Evol_trace}) and (\ref{princ_PI}) are also solutions to the full set of Einstein's equations provided that (\ref{ort_const}) holds on the initial data surface $\Sigma_0$. 
\end{corollary}

Note that this result may be consider{ed} as a generalization of the mixed initial value problem applied, e.g.~by Choptuik \cite{choptuik} (see also section 2 of \cite{racz_KoVeFi} for a detailed explanation) in investigating the critical phenomenon in spherically symmetric spacetimes. It is worth emphasizing that the present setup allows more general slicings and even if considerations were restricted to spherically symmetric configurations the `radial coordinate' $\rho$ would not need to coincide with the area radius applied in \cite{choptuik}. 

\section{Final remarks}\label{final}
\setcounter{equation}{0}

In this paper the Cauchy problem in Einstein's theory of gravity was considered. It was assumed that the spacetime manifold can be foliated by a two-parameter family of homologous two-surfaces. By providing a constructional definition of the conformal structure of the foliat{ing} two-surfaces and applying the canonical form of the spacetime metric a new gauge fixing had been introduced. Remarkably the $1+3$ Hamiltonian and momentum constraints which are purely elliptic in the conventional conformal approach \cite{wheeler, york, york0, fisher} could be put into the form of a coupled parabolic-hyperbolic system. This latter system can conveniently be solved as a (boundary-)initial value problem thereby the new method appears to be more favorable than the conventional one. Whenever the two-surfaces foliating the spacetime are compact orientable with no boundary the pertinent initial value problem has been found at least in certain particular cases to be well-posed. 

\medskip

The full $1+3$ reduced Einstein's equations are also determined. 
By combining the $1+3$ momentum constraint with the reduced system of the secondary $1+2$ decomposition a mixed hyperbolic-hyperbolic system is formed. It is shown that solutions to this mixed hyperbolic-hyperbolic system are also solutions to the full set of Einstein's equations provided that the $1+3$ Hamiltonian constraint holds on the initial data surface $\Sigma_0$ and the $1+2$ Hamiltonian and momentum type expressions vanish on a world-tube yielded by the Lie transport of one of the two-surfaces foliating $\Sigma_0$ along the time evolution vector field.\,\footnote{Note that this result provides an immediate {inspiration to} revisit the conventional initial-boundary value problem which will be done in \cite{racz_ibvp}.} Whenever the foliating two-surfaces are compact without boundary in the spacetime and a regular origin exists on the time-slices, in addition to the mixed hyperbolic-hyperbolic system, it suffices to solve the $1+3$ Hamiltonian constraint on the initial data surface.

\medskip

The introduced new analytic setup appears to be {very promising} concerning the solubility of the constraint and evolution equations. Nevertheless, we have to admit that the mere introduction of the proposed method opened a lot more new issues than answers yet. Without aiming to give a complete list of the open problems let us mention only some immediate ones:

\begin{itemize}
\item Under which conditions the coupled parabolic-hyperbolic system (\ref{scal_constr_ell}) and (\ref{constr_hyp}) possesses well-posed initial value formulation provided that the foliating two-surfaces are of arbitrary genus compact orientable two-manifolds with no boundary in the spacetime? 
\item Assuming that satisfactory well-posedness results can be deduced in the above setup. In what extent could the corresponding results be generalized to foliations with two-surfaces which are either compact with boundary or non-compact but infinite? Supposedly, in the latter case instead of the boundary conditions some sort of asymptotic fall off conditions have to be applied. 
\item Is there any way to economize the generality of the introduced analytic setup in providing some new insight concerning the positive mass theorem or the Penrose inequality in the fashion  attempt to be solved in \cite{gerochEn, jang,jang_wald,bartnik,schwein1,schwein2}? 
\item {The mixed hyperbolic-hyperbolic evolutionary system is comprised by two systems one of which is first order symmetric hyperbolic system whereas the other is second order strongly hyperbolic one. As these separate subsystems are well-posed the combined system is also expected to be so. Nevertheless, it has to be checked whether this expectation comes through.}
\item Would it be possible to apply the new analytic setup in numerical simulations? It appears that by making use of GridRipper \cite{gridripper}---based on the use of foliations by topological two-spheres---the coupled parabolic-hyperbolic system (\ref{scal_constr_ell}) and (\ref{constr_hyp}) could be put into a set of coupled ordinary differential equations for the coefficients yielded by the multipole expansion of the basic variables.  
\end{itemize}

\medskip

It is worth emphasizing that in deriving our results considerations were restricted exclusively to the geometric part, i.e.~to Einstein's equations. If matter fields are involved the pertinent Euler-Lagrange equations have also to be taken into account. In general, even if the matter field variables have certain gauge freedom those are considered to be more familiar, and thereby easier to be handled than the one generated by diffeomorphism invariance of Einstein's theory of gravity. Nevertheless, one has to keep in mind that whenever matter fields are included careful case by case investigations have to complement the analysis carried out in this paper.

\medskip

By virtue of the constraint and evolution equations (\ref{par_const2}), (\ref{ort_const2}), (\ref{scal_const}) and (\ref{Evol_nn}), (\ref{Evol_pn}), (\ref{Evol_trace}),  (\ref{princ_PI}) it gets immediately transparent that they look as if they were given in terms of fields---along with their transversal `$\tau$'- and `$\rho$'-derivatives---defined exclusively on the two-surfaces $\mycal{S}_{\tau,\rho}$ foliating the spacetime. Therefore, by suppressing the spacetime origin of the applied variables, one could also interpret these equations as if they governed the evolution of quantities living on generic two-surfaces $\mycal{S}_{\tau,\rho}$. These observations naturally lead to the concept of two-dimensional `geometrodynamics' which---in its spirit (though not in other details)---is the same as the concept of `geometrodynamics' proposed by Wheeler \cite{wheeler} (see also \cite{fisher}) in case of $1+3$ decompositions. Accordingly, the conventional Cauchy problem may also be viewed as a two-surface based 
`geometrodynamics'.

\medskip

By inspecting the freely specifiable variables on the initial data surface $\Sigma_0$ we may also acquire some hints about the geometric degrees of freedom of Einstein's theory. Recall first that on $\Sigma_0$ the induced metric $h_{ij}$ can be given in terms of $\hat N$, $\hat N^{A}, \Omega, \gamma_{AB}$ as in (\ref{hijind}). In general, the lapse and shift $\hat N$ and $\hat N^{A}$ are supposed merely to fix the foliation of $\Sigma_0$ by the two surfaces $\mycal{S}_{\rho}$. Notice also that we have a restricted freedom in choosing $\hat N$, $\hat N^{A}$ and $\Omega$ since merely half of their initial data {on $\Sigma_0$} can be chosen at our will as $\hat N, \mycal{L}_n\Omega$ and $\mycal{L}_n \hat N^{A}$ are subject to (\ref{scal_constr_ell}) and (\ref{constr_hyp}). As opposed to these variables complete freedom is available in choosing the conformal structure $\gamma_{AB}$ and $\mycal{L}_n\,\gamma_{AB}$ throughout $\Sigma_0$.\,\footnote{It was already suggested by discussions in \cite{diverno1} that 
`these four pieces 
of information embody the two gravitational degrees of freedom' for the Cauchy problem considered in this paper.} Thereby $\gamma_{AB}$, possessing only two algebraically independent components, appears to represent {the}
gravitational degrees of freedom. A more adequate determination of the gravitational degrees of freedom can be provided by dynamical investigations (for more details see \cite{racz_tdfd}).

\section*{Acknowledgments}

The author is grateful to Lars Anderson, Bobby Beig, Ingemar Bengtsson, Bob Wald {and the unknown referees} for helpful comments and suggestions. 
This research was supported by the European Union and the State of Hungary, co-financed by the European Social Fund in the framework of T\'AMOP-4.2.4.A/2-11/1-2012-0001 ``National Excellence Program''. The author is also grateful to the Albert Einstein Institute in Golm, Germany for its kind hospitality where parts of the reported results were derived. 

\appendix
\section{Appendix: The source terms}\label{Appendix A}
\renewcommand{\theequation}{A.\arabic{equation}}
\setcounter{equation}{0}

To keep the expressions below to be reasonably compact in most of the cases only the second order derivatives of the variables $\Omega, \gamma_{ij}, \hat N, \hat N^i;\nu$ will be spelled out explicitly.

\medskip

Consider first $\hat{\mathfrak{e}}=\hat n^i \hat n^j\,{}^{{}^{(3)}}\hskip-1mm \mycal{G}_{ij}$. Recall first that in virtue of (\ref{int_evol_1+3_ch}) 
\begin{eqnarray}
&& \hskip-0.8cm\hat{\mathfrak{e}}  = \mathfrak{S}_{ij}\,\hat n^i \hat n^j + \left\{\phantom{\tfrac12}\hskip-0.4cm-(\mycal{L}_n K_{ij}) \,\hat n^i \hat n^j- ({K^{l}}_{l}) \,\boldsymbol\kappa + 2\,(\boldsymbol\kappa^2+{\rm\bf k}^{l}{\rm\bf k}{}_{l}) +{N}^{-1}\,\hat n^i \hat n^j\,(D_i D_j N) \right.  \nonumber \\ && \hskip-1.1cm \phantom{\hat{\mathfrak{e}}= = \mathfrak{S}_{ij}\,\hat n^i \hat n^j -\epsilon\left\{ \right. }\left. + \mycal{L}_n ({K^l}_{l})+\tfrac12\, ({K^{l}}_{l})^2 +\tfrac12\,K_{kl}{K^{kl}} - {N}^{-1}\,D^l D_l N \right\} \,,
\end{eqnarray}
where 
\begin{equation}
K_{il}{K^{l}}_{j}\,\hat n^i \hat n^j=\hat n^i\,K_{il}\,[\hat\gamma{}^{kl}+\hat n^k \hat n^l]\,K_{lj}\, \hat n^j=\boldsymbol\kappa^2+2\,{\rm\bf k}^{l}{\rm\bf k}{}_{l}
\end{equation}
was applied. 

\medskip

Notice also that 
\begin{equation}\label{DDN0}
D^l D_l N -\hat n^i \hat n^j \,D_i D_j N= [h^{kl}-\hat n^k \hat n^l]\,D_k D_l N = \hat\gamma{}^{kl}\,D_k D_l N = ({{\hat K}{}^l}_{l}) \mycal{L}_{\hat n} N + \hat D^l \hat D_l N \,,
\end{equation}
where in the last step the contraction of $\hat\gamma{}^{kl}$ and the decomposition\,\footnote{(\ref{DDN}) may be verified by using (A.4) of \cite{racz_geom_det} with the replacements $\nabla_a\rightarrow D_a$ and $L_a \rightarrow D_a N$. }
\begin{eqnarray}\label{DDN}
&& \hskip-0.8cm D_k D_l N = [\hat D_k\,(\mycal{L}_{\hat n} N) + \hat n_k\,(\mycal{L}^2_{\hat n} N) ]\,\hat n_l + (\mycal{L}_{\hat n} N)\,[\hat K{}_{kl}+ \hat n_k\, \dot{\hat n}{}_l] + \hat D_k \hat D_l N  \\  
&& \hskip-0.8cm  \phantom{D_k D_l N =} -\hat n_k\, \hat n{}_l\,[\dot{\hat n}{}^p\,\hat D_p N ] +\hat n_k\,\mycal{L}_{\hat n}(\hat D_l N)-\hat n_k\, \hat K{}^p{}_l\,(\hat D_p N) - \hat n_l\, \hat K{}^p{}_k\,(\hat D_p N) \,, \nonumber
\end{eqnarray}
was applied.

\medskip

In virtue of (\ref{loc_expr41}), (\ref{loc_expr42}), (\ref{dec_2}) and (\ref{loc_expr_liek1}) we have
\begin{equation}
(\mycal{L}_n K_{ij}) \,\hat n^i \hat n^j=\mycal{L}^2_n (\ln \hat N) + 2\,\boldsymbol\kappa^2 +4\,{\rm\bf k}^{l}{\rm\bf k}{}_{l}\,.
\end{equation}
Then by combining all the above relations with (\ref{dec_2}) and 
\begin{equation}
\mycal{L}_n ({K^l}_{l})=\mycal{L}^2_n (\ln \hat N)+\mycal{L}^2_n (\ln \Omega^2)
\end{equation}
we get 
\begin{eqnarray}\label{ehat}
&& \hat{\mathfrak{e}}=\mathfrak{S}_{ij}\,\hat n^i \hat n^j + \mycal{L}^2_n (\ln \Omega^2) -{\rm\bf k}^{l}{\rm\bf k}{}_{l} + \tfrac12\,({\rm\bf K}^{l}{}_l)^2  + \tfrac12\,{\rm\bf K}_{kl}{\rm\bf K}^{kl}  \\ && \phantom{\hat{\mathfrak{e}}=\mathfrak{S}_{ij}\,\hat n^i \hat n^j + \phantom{\frac{\epsilon}{N}}\hskip-0.5cm \mycal{L}^2_n (\ln \Omega^2)}- [\hat D{}^l \hat D{}_l \ln N + (\hat D{}^l \ln N) (\hat D{}_l \ln N)  + ({{\hat K}{}^l}_{l}) \mycal{L}_{\hat n} \ln N ] \,.\nonumber
\end{eqnarray}

\medskip

In determining $\hat{\mathfrak{p}}_{i}={{\hat \gamma}^{k}}{}_{i}\,\hat n^l\, {}^{{}^{(3)}}\hskip-1mm \mycal{G}_{kl}$ we may proceed in an analogous way. Notice that 
\begin{eqnarray}\label{phat0}
&&  \hskip-0.6cm \hat{\mathfrak{p}}_{i}= {{\hat \gamma}^{k}}{}_{i}\hat n^l\, \mathfrak{S}_{kl} -\left\{ (\mycal{L}_n K_{kl})\,{{\hat \gamma}^{k}}{}_{i}\hat n^l + ({K^{l}}_{l}) \,{\rm\bf k}{}_{i} - 2\,(\boldsymbol\kappa\,{\rm\bf k}{}_{i}+{\rm\bf k}^{l}{\rm\bf K}{}_{il}) - {N}^{-1}\,(D{}_k  D{}_l N)\,{{\hat \gamma}^{k}}{}_{i}\hat n^l  \right\}  \nonumber \\
\end{eqnarray}
where the relation
\begin{equation}
{{\hat \gamma}^{k}}{}_{i}\,\hat n^l\,K_{kp}{K^{p}}_{l}={{\hat \gamma}^{k}}{}_{i}\,K_{kp}\,[\hat\gamma{}^{pq}+\hat n^p \hat n^q]\,K_{ql}\, \hat n^l={\rm\bf k}^{l}{\rm\bf K}{}_{il}+\boldsymbol\kappa\,{\rm\bf k}{}_{i}
\end{equation}
was used. 

\medskip

Applying (\ref{DDN}) we get 
\begin{eqnarray}\label{DDNhat}
&& {N}^{-1} {{\hat \gamma}^{k}}{}_{i}\,\hat n^l\,(D_k D_l N)= {N}^{-1} \left[\hat D_i[\mycal{L}_{\hat n} N ] -  {{\hat K}{}^{k}}{}_{i}\,(\hat D_k N) \right]  \\ && \phantom{{N}^{-1} {{\hat \gamma}^{k}}{}_{i}\,\hat n^l\,(D_k D_l N)} = \mycal{L}_{\hat n} (\ln N)\,\hat D_i[\ln N] + \hat D_i [\mycal{L}_{\hat n} (\ln N)] -  {{\hat K}{}^{k}}{}_{i}(\hat D_k \ln N)\,, \nonumber
\end{eqnarray}
which, along with  (\ref{loc_expr_liek2}), (\ref{taurhoSection1}), (\ref{loc_expr_liek2}) and some algebra, yields 
\begin{eqnarray}\label{phat}
&& \hskip-1.6cm \hat{\mathfrak{p}}_{i}= {{\hat \gamma}^{k}}{}_{i}\,\hat n^l\, \mathfrak{S}_{kl} - {(2\hat N)}^{-1}\Omega^2
\left[\gamma_{kl}(\mycal{L}^2_{n}\hat N^l)+2\,(\mycal{L}_{n} \gamma_{kl})(\mycal{L}_{n}\hat N^l) + 2\,\gamma_{kl}(\mycal{L}_{n}\hat N^l)(\mycal{L}_n \ln \Omega^2) \phantom{\tfrac12}\hskip-.3cm\right]
 \nonumber \\ && \hskip-1.6cm \phantom{\hat{\mathfrak{p}}_{i}=} 
+ \boldsymbol\kappa\,{\rm\bf k}{}_{i}+ 2\,\interior{\rm\bf K}{}_{il}{\rm\bf k}^{l}  
+\left[\mycal{L}_{\hat n} (\ln N)\,[\hat D_i(\ln N)] + \hat D_i [\mycal{L}_{\hat n} (\ln N)] - {{\hat K}{}^{k}}{}_{i}(\hat D_k \ln N)\right]\hskip-0.1cm.
\end{eqnarray} 

\medskip

In determining $\hat \gamma^{kl}\,\hat{\mathfrak{S}}_{kl}=\hat \gamma^{kl}\,({{\hat \gamma}^{p}}{}_{k} {{\hat \gamma}^{q}}{}_{l}\,{}^{{}^{(3)}}\hskip-1mm \mycal{G}_{pq})=\hat \gamma^{kl}\,{}^{{}^{(3)}}\hskip-1mm \mycal{G}_{kl}$ recall first that by (\ref{int_evol_1+3_ch}) 
\begin{eqnarray}\label{ghat_sigmahat}
\hat \gamma^{kl}\,\hat{\mathfrak{S}}_{kl} {} & \hskip-2mm = & \hskip-1mm {} \hat \gamma^{kl}\, {\mathfrak{S}}_{ij} + \left\{ -\hat \gamma^{kl}\,(\mycal{L}_n K_{kl}) - ({K^{l}}_{l})\, {\rm\bf K}{}^{k}{}_k + 2\, {K}_{kp}\,{K^{p}}_{l}\,\hat \gamma^{kl} + {N}^{-1}\, \hat \gamma^{kl}\,D_kD_l N \right. \nonumber \\ && \hskip23mm \left.  + 2\,\mycal{L}_n ({K^{l}}_{l}) + ({K^{l}}_{l})^2 + {K}_{kl}\,{K^{kl}} - 2\,{\hat N}^{-1}\,D^l D_l N \right\}\,.
\end{eqnarray}
By combining then 
\begin{equation}\label{extKD2}
\mycal{L}^2_{n} \hat \gamma_{ij}= \mycal{L}^2_{n} (\ln  \Omega^2)\,\hat \gamma_{ij}+  \left(\mycal{L}_{n} (\ln  \Omega^2)\right)^2 \,\hat \gamma_{ij} + 2\,\mycal{L}_{n} (\ln  \Omega^2)\,\Omega^2\, \mycal{L}_{n}\gamma_{ij} + \Omega^2\, \mycal{L}^2_{n}\gamma_{ij} 
\end{equation}
with (\ref{loc_expr_liek3}) we get that 
\begin{eqnarray}\label{ghat_sigmahat_l1}
&& \hat \gamma^{kl}\,(\mycal{L}_n K_{kl}) = \tfrac12\,\hat \gamma^{kl}\mycal{L}^2_{n} \hat \gamma_{kl} = \mycal{L}^2_{n} (\ln  \Omega^2) + [\mycal{L}_{n} (\ln  \Omega^2)]^2 +\tfrac12\,\gamma^{kp} \gamma^{lq} (\mycal{L}_{n}\gamma_{kl}) (\mycal{L}_{n}\gamma_{pq}) \nonumber \\ && \phantom{\hat \gamma^{kl}\,(\mycal{L}_n K_{kl}) = \tfrac12\,\hat \gamma^{kl}\mycal{L}^2_{n} \hat \gamma_{kl}} = \mycal{L}^2_{n} (\ln  \Omega^2) + 2\,{\rm\bf K}{}_{kl}{\rm\bf K}{}^{kl} \,,
\end{eqnarray}
where in the second step (\ref{xi}), with $\eta^a=\tau^a$, i.e.~$\gamma^{kl}(\mycal{L}_{n}\gamma_{kl})\equiv 0$, and 
\begin{equation}\label{lll}
\gamma^{kl}(\mycal{L}^2_{n}\gamma_{kl})=\mycal{L}_{n}[\gamma^{kl}(\mycal{L}_{n}\gamma_{kl})]-(\mycal{L}_{n}\gamma^{kl})(\mycal{L}_{n}\gamma_{kl})= \gamma^{kp} \gamma^{lq} (\mycal{L}_{n}\gamma_{kl}) (\mycal{L}_{n}\gamma_{pq})\,,
\end{equation}
while in the last step (\ref{interiors}) and (\ref{dec_4}) were applied. 

\medskip

Similarly, we have 
\begin{eqnarray}\label{6.21}
2\, {K}_{kp}\,{K^{p}}_{l}\,\hat \gamma^{kl}= 2\, {K}_{kp}\,{K}_{ql}\,\hat \gamma^{kl}\, [\hat \gamma^{pq}+\hat n^{p}\hat n^{q}]= 2\,[{\rm\bf K}{}_{kl}{\rm\bf K}{}^{kl}+{\rm\bf k}{}_{l}{\rm\bf k}{}^{l}]
\end{eqnarray}
and 
\begin{eqnarray}
D^l D_l N-\hat \gamma^{kl}\,D_kD_l N {} & \hskip-2mm = & \hskip-1mm {} [\gamma^{kl}+\hat n^{k}\hat n^{l}] (D_k D_l N) -\hat \gamma^{kl}\,D_kD_l N = \hat n^{k}\hat n^{l}\, D_k D_l N \nonumber \\ && \hskip-2mm 
=\mycal{L}^2_{\hat n} N + \hat D^{k} (\ln \hat N)\hat D_{k} N\,,
\end{eqnarray}
where in the last step (\ref{DDN}), along with $\dot{\hat n}^{k}= - \hat D^{k} (\ln \hat N)$, was used. In addition, by (\ref{DDN0}), we also have 
\begin{equation}\label{6.23}
D^l D_l N = [\gamma^{kl}+\hat n^{k}\hat n^{l}] (D_k D_l N) =  ({\hat K}{}^{l}{}_{l})\,\mycal{L}_{\hat n} N +  \hat D^l \hat D_l N + \hat n^{k} \hat n^{l} D_k D_l N \,.
\end{equation}
Thus, by combining (\ref{ghat_sigmahat_l1}), (\ref{6.21})-(\ref{6.23}), and using (\ref{loc_expr41})-(\ref{dec_2}) and (\ref{trhatK}), along with some algebra, we get 
\begin{eqnarray}\label{1_ghat_sigmahat}
\hat \gamma^{kl}\,\hat{\mathfrak{S}}_{kl} {} & \hskip-2mm = & \hskip-1mm {} \hat \gamma^{kl}\, {\mathfrak{S}}_{kl} +
2\, \mycal{L}^2_n (\ln \hat N) + \mycal{L}^2_n (\ln \Omega^2) + 2\,\boldsymbol\kappa^2 + \boldsymbol\kappa\,{\rm\bf K}{}^{l}{}_l 
+ 4\,{\rm\bf k}{}_{l}\,{\rm\bf k}{}^{l}+ {\rm\bf K}{}_{kl}\,{\rm\bf K}{}^{kl}  \nonumber \\ &&  \phantom{\hat \gamma^{kl}\, {\mathfrak{S}}_{ij}}\hskip-1mm - \left[ 2\, \mycal{L}^2_{\hat n} (\ln  N) +2\, [\mycal{L}_{\hat n} (\ln  N)]^2 + 2\,(\hat D^{k} \ln \hat N)(\hat D_{k} \ln N) \right. \nonumber \\ && \left. \phantom{\hat \gamma^{kl}\, {\mathfrak{S}}_{ij} - } \hskip1mm  + ({\hat K}{}^{l}{}_{l})\,\mycal{L}_{\hat n} (\ln N) 
+\hat D^l \hat D_l \ln N +(\hat D^l \ln N)(\hat D_l \ln N)
\right]  \,.
\end{eqnarray}

\medskip

Before determining $\Pi^{kl}{}_{ij}\hat{\mathfrak{S}}_{kl}=\Pi^{kl}{}_{ij}\,[{{\hat \gamma}^{p}}{}_{k} {{\hat \gamma}^{q}}{}_{l}\,{}^{{}^{(3)}}\hskip-1mm \mycal{G}_{pq}]=\Pi^{kl}{}_{ij}\,{}^{{}^{(3)}}\hskip-1mm \mycal{G}_{kl}$ note first that by (\ref{PI_def}), (\ref{PI_negyzet}) and (\ref{PI2}) the relation 
\begin{equation}\label{PI1}
\Pi^{kl}{}_{ij}\,\hat \gamma^p{}_k\hat \gamma^q{}_l=\hat \gamma^p{}_i\hat \gamma^q{}_j-\tfrac12\,\hat \gamma_{ij}\,\hat \gamma^{pq}=\Pi^{pq}{}_{ij}
\end{equation}
holds.

\medskip

In addition, by making use of (\ref{int_evol_1+3_ch}), along with (\ref{PI1}) and (\ref{PI2}), we have that 
\begin{eqnarray}\label{PI_sigmahat}
\Pi^{kl}{}_{ij}\,\hat{\mathfrak{S}}_{kl} {} & \hskip-2mm = & \hskip-1mm {} \Pi^{kl}{}_{ij}\, {{\mathfrak{S}}_{kl}} - \Pi^{kl}{}_{ij}\, (\mycal{L}_n K_{kl}) - ({K^{l}}_{l})\, \interior{\rm\bf K}{}_{ij} + 2\,\Pi^{kl}{}_{ij}\,( {K}_{kp}\,{K^{p}}_{l})  \\ && \hskip-22mm \phantom{\Pi^{kl}{}_{ij}\,\hat{\mathfrak{S}}_{kl} {} =\Pi^{kl}{}_{ij}\, {\mathfrak{S}}_{ij} - \Pi^{kl}{}_{ij}\, (\mycal{L}_n K_{kl}) }  + \Pi^{kl}{}_{ij}\, \left[D_kD_l \ln N+ (D_l \ln N) (D_l \ln N)\right] \,.\nonumber
\end{eqnarray}
Note that by combining (\ref{extKD2}), (\ref{lll}),  (\ref{loc_expr_liek3}) and (\ref{PI2}) we get then 
\begin{eqnarray}\label{PI_sigmahat_l1}
\Pi^{kl}{}_{ij}\,\mycal{L}_{n} K_{kl} {} \hskip-2mm & = & \hskip-2mm {} \tfrac12\,{\Omega^2}\left[\Pi^{kl}{}_{ij}\,\mycal{L}^2_{n}\gamma_{kl}  + \mycal{L}_{n} (\ln  \Omega^2)\,\Pi^{kl}{}_{ij}\,\mycal{L}_{n}\gamma_{kl}\right]  \\ {}  \hskip-2mm & \hskip-12mm = & \hskip-2mm {} \hskip-6mm\tfrac12\,{\Omega^2}\left[\gamma^k{}_i \gamma^l{}_j\left[\mycal{L}^2_{n}\gamma_{kl}  + 2\,\mycal{L}_{n} (\ln  \Omega^2)\,\mycal{L}_{n}\gamma_{kl}\right] - \tfrac12\,\gamma_{ij}\left[ \gamma^{kp} \gamma^{lq} (\mycal{L}_{n}\gamma_{kl}) (\mycal{L}_{n}\gamma_{pq})\right]\right]\,.\nonumber 
\end{eqnarray}

\medskip

In evaluating the other second order term in (\ref{PI_sigmahat}) note that 
\begin{eqnarray}\label{PI_DDN}
\Pi^{kl}{}_{ij}\,(D_kD_l \ln N) {} \hskip-2mm & = & \hskip-2mm {} \hat \gamma^k{}_i\hat \gamma^l{}_j(D_kD_l \ln N) - \tfrac12\,\hat \gamma_{ij}[\hat \gamma^{kl}(D_kD_l \ln N)] \nonumber  \\ {} \hskip-2mm & = & \hskip-2mm {} \hat D_i \hat D_j \ln N - \tfrac12\,\hat \gamma_{ij} \,(\hat D^l \hat D_l \ln N) + \mycal{L}_{n}(\ln  N) \,\interior{\hat K}_{ij}  
\end{eqnarray}
where in the second line (\ref{DDN}) was applied.

\medskip

Finally by combining (\ref{PI_sigmahat}), (\ref{PI_sigmahat_l1}) and (\ref{PI_DDN}) we get
\begin{eqnarray}\label{PI_sigmahat1}
\Pi^{kl}{}_{ij}\,\hat{\mathfrak{S}}_{kl} {} & \hskip-2mm = & \hskip-3mm {} -\tfrac12\,{\Omega^2}\left[\gamma^k{}_i \gamma^l{}_j\left[\mycal{L}^2_{n}\gamma_{kl}  + 2\,\mycal{L}_{n} (\ln  \Omega^2)\,\mycal{L}_{n}\gamma_{kl}\right] - \tfrac12\, \gamma_{ij}\left[ \gamma^{kp} \gamma^{lq} (\mycal{L}_{n}\gamma_{kl}) (\mycal{L}_{n}\gamma_{pq})\right]\right] \nonumber \\ && \hskip+2mm + \hat D_i \hat D_j \ln N - \tfrac12\,\hat \gamma_{ij} \,(\hat D^l \hat D_l \ln N) + \mycal{L}_{n}(\ln  N) \,\interior{\hat K}_{ij} \nonumber \\ && \hskip+2mm 
+\Pi^{kl}{}_{ij} \left\{ {\mathfrak{S}}_{kl} - ({K^{p}}_{p})\, \interior{\rm\bf K}{}_{kl} + 2\,{K}_{kp}\,{K^{p}}_{l} + (D_k \ln N) (D_l \ln N) \phantom{\frac{\epsilon}{\epsilon}} \hskip-3mm\right\} \,.
\end{eqnarray}


\end{document}